\documentclass[11pt,a4paper]{article}
\usepackage[utf8]{inputenc}

\usepackage{jheppub}

\RequirePackage{ifpdf} 
\usepackage{amsmath} 
\usepackage{mathtools}
\usepackage{pstricks}
\usepackage[final]{pdfpages} 
\usepackage{ifpdf} 
\usepackage{slashed}
\usepackage{hyperref}
\usepackage{color} 
\usepackage{graphics}
\usepackage{lineno}

\usepackage{etoolbox} % Load before 'breqn' package to make 'lineno' work
\usepackage{fixmath}

\usepackage{amsfonts}

\usepackage{multirow}
\usepackage{epstopdf}

\usepackage[normalem]{ulem}

\usepackage{tikz}
\usetikzlibrary{positioning,arrows}
\usetikzlibrary{decorations.pathmorphing}
\usetikzlibrary{decorations.markings}
\usetikzlibrary{shapes.geometric}
\tikzset{
    vector/.style={decorate, decoration={snake}, draw},
    provector/.style={decorate, decoration={snake,amplitude=2.5pt}, draw},
    antivector/.style={decorate, decoration={snake,amplitude=-2.5pt}, draw},
    fermion/.style={draw=black,
      postaction={decorate},decoration={markings,mark=at position .55
        with {\arrow[draw=black]{>}}}}, 
    fermionbar/.style={draw=black, postaction={decorate},
                       decoration={markings,mark=at position .55 with 
{\arrow[draw=black]{<}}}},
    fermionnoarrow/.style={draw=black},
    gluon/.style={decorate, draw=black,decoration={coil,amplitude=4pt, segment 
length=4pt}},
    scalar/.style={dashed,draw=black,
      postaction={decorate},decoration={markings,mark=at position .55
        with {\arrow[draw=black]{>}}}}, 
    scalarbar/.style={dashed,draw=black,
      postaction={decorate},decoration={markings,mark=at position .55
        with {\arrow[draw=black]{<}}}}, 
    scalarnoarrow/.style={dashed,draw=black},
    electron/.style={draw=black,
      postaction={decorate},decoration={markings,mark=at position .55
        with {\arrow[draw=black]{>}}}}, 
    bigvector/.style={decorate, decoration={snake,amplitude=4pt}, draw},
} 
\allowdisplaybreaks[4]
\newcommand{\nnn}{\nonumber\\}

\preprint{IMSc/2017/08/06}

\title{\boldmath Two loop QCD corrections for the process Pseudo-scalar Higgs $\rightarrow 3$ partons}
\author{Pulak Banerjee$^{a,b}$, Prasanna K.\
  Dhani$^{a,b}$ and V. Ravindran$^{a,b}$} 
  
  \affiliation[a]{The Institute of Mathematical Sciences, Taramani,
  Chennai 600113, India}
  \affiliation[b]{Homi Bhaba National Institute,
  Training School Complex, Anushakti Nagar, \\Mumbai 400085, India }

\emailAdd{bpulak@imsc.res.in}
\emailAdd{prasannakd@imsc.res.in}
\emailAdd{ravindra@imsc.res.in}
\abstract{
We present virtual contributions up to two loop level in perturbative Quantum Chromodynamic (QCD) 
to the decay of pseudo-scalar Higgs boson ($A$) to three gluons ($g$) and 
also to quark ($q$), anti-quark ($\overline q$) and a gluon. With appropriate crossing, they are well suited for  
predicting the differential distribution of $A$ in association with a jet in hadron colliders up to next-to-next-to-leading order (NNLO) in strong coupling constant and also for the subsequent decay of $A$ to hadrons.
We use effective field theory approach to integrate out the top quarks in the heavy top limit.   
The resulting theory involves two pseudo-scalar composite operators describing the interaction of $A$ with gluons as well as  with 
quark and anti-quark.  
We perform our computation in dimensional regularisation and use minimal subtraction ($\overline{MS}$) scheme
to renormalise strong coupling constant as well as the composite operators. The ultraviolet (UV) finite amplitudes contain 
infrared (IR) divergences that are found to be in agreement with the predictions by Catani.  For both the amplitudes namely
$A \rightarrow g g g$ and $A \rightarrow q \overline q g$, the leading transcendental terms 
at one and two loops are found to be identical to those in a three point form factor (FF) of the half-BPS operator
in ${\cal N}=4$ Supersymmetric Yang Mills (SYM) theory when the QCD color factors are adjusted in a specific way. 
We present our results in terms of harmonic polylogs well suited 
for further numerical study.}
\begin{document} 

\maketitle
% \flushbottom

%************
% Definition
%************

\def\D{{\cal D}}
\def\DD{\overline{\cal D}}
\def\g{\overline{\cal G}}
\def\gm{\gamma}
\def\M{{\cal M}}
\def\uM{{\hat{\cal M}}}
\def\ep{\epsilon}
\def\epm1{\frac{1}{\epsilon}}
\def\epm2{\frac{1}{\epsilon^{2}}}
\def\epm3{\frac{1}{\epsilon^{3}}}
\def\epm4{\frac{1}{\epsilon^{4}}}
\def\unM{\hat{\cal M}}
\def\ashat{\hat{a}_{s}}
\def\asmur{a_{s}^{2}(\mu_{R}^{2})}
\def\sigbar{{{\overline {\sigma}}}\left(a_{s}(\mu_{R}^{2}), L\left(\mu_{R}^{2}, 
m_{H}^{2}\right)\right)}
\def\sigbarn{{{{\overline \sigma}}_{n}\left(a_{s}(\mu_{R}^{2}) 
L\left(\mu_{R}^{2}, m_{H}^{2}\right)\right)}}
\def\unas{ \left( \frac{\hat{a}_s}{\mu_0^{\epsilon}} S_{\epsilon} \right) }
\def\rnM{{\cal M}}
\def\bt{\beta}
\def\cD{{\cal D}}
\def\cC{{\cal C}}
\def\ca{\text{\tiny C}_\text{\tiny A}}
\def\cf{\text{\tiny C}_\text{\tiny F}}
\def\ct{{\red []}}
\def\sv{\text{SV}}
\def\murOmu{\left( \frac{\mu_{R}^{2}}{\mu^{2}} \right)}
\def\bb{b{\bar{b}}}
\def\bt0{\beta_{0}}
\def\bt1{\beta_{1}}
\def\bt2{\beta_{2}}
\def\bt3{\beta_{3}}
\def\gm0{\gamma_{0}}
\def\gm1{\gamma_{1}}
\def\gm2{\gamma_{2}}
\def\gm3{\gamma_{3}}
\def\nn{\nonumber}
\def\l{\left}
\def\r{\right}
%%%%

%\newcommand{\dis}[1]{\color{blue}\mathbold{#1}}
\newcommand{\dis}{}
\newcommand{\overbar}[1]{mkern-1.5mu\overline{\mkern-1.5mu#1\mkern-1.5mu}\mkern
1.5mu}

\section{Introduction}
\label{sec:intro}
The discovery of a scalar particle in 2012 at the Large 
Hadron Collider (LHC) \cite{Aad:2012tfa,Chatrchyan:2012xdj}  
is the milestone in high energy physics. The detailed study on 
the nature of this particle and its couplings to other Standard Model (SM) particles is underway. 
There are already several compelling evidences pointing to the fact that 
the discovered boson is none other than the Higgs boson of the SM. 
Although the SM is remarkably successful in explaining most of the 
observed phenomena at the subatomic level, it has many shortfalls.  For example, 
it does not have dark matter candidate to explain relic abundance in the early universe, similarly
the observed baryon asymmetry and the phenomena of neutrino oscillations.  
There exist several beyond the SM (BSM) scenarios that address these issues and provide
plausible explanations.  Often they have larger symmetry and contain more particles than
the SM.  Most of the BSMs contain the scalar sector with additional scalars providing the scope
for rich phenomenology at the LHC.
The precise measurements of the coupling of the Higgs boson with other SM 
particles as well as the dedicated direct searches of new scalars at the LHC
can constrain the parameters of these scalar sectors in the BSM models. 
For example, in the Minimal Supersymmetric Standard Model (MSSM)
which addresses some of the shortfalls of the SM contains in addition to other 
fields, two isospin doublets of Higgs field 
~\cite{Fayet:1974pd,Fayet:1976et,Fayet:1977yc,Dimopoulos:1981zb,Sakai:1981gr,
Inoue:1982pi,Inoue:1982ej,Inoue01021984}.  These two doublets preserve 
the analyticity of the scalar potential
and also maintain the anomaly cancellation. They are responsible for 
masses of up and down type fermions. After spontaneous symmetry breaking, scalar sector
contains three neutral ($h,H,A$) and two 
charged ($H^{\pm}$) Higgs bosons, where $h$ and $H$ are CP-even scalars while $A$ is CP-odd 
pseudo-scalar.  The upper bound on the mass of the light scalar Higgs $h$ 
can be predicted within the theory as the self couplings of the Higgs fields are fixed in
terms of the gauge couplings.        

Since the coupling of $A$ with the fermions is appreciable 
in the small and moderate $\tan \beta$, 
the ratio of vacuum expectation values $v_i, i = 1, 2$ of the Higgs doublets, it can be searched
at the LHC.  Note that large gluon flux at the LHC can also boost the cross section.
While, the searches for $A$ have been underway at the LHC,
predictions for observables involving $A$, in the theoretical side, are already available.  
Like the production of CP scalar Higgs boson, the leading order predictions for $A$ severely suffer
from theoretical uncertainties resulting from renormalisation ($\mu_R$) and 
factorisation ($\mu_F$) scales as well as large higher order radiative corrections.     
These scales enter at leading order through the renormalised 
strong coupling constant $g_s(\mu_R)$ and the parton distribution functions defined
at the factorisation scale $\mu_F$.  This necessitates to go beyond
leading order in perturbation theory.  In \cite{Kauffman:1993nv,Djouadi:1993ji, Spira:1993bb, Spira:1995rr}
the next-to-leading order (NLO) perturbative QCD corrections to inclusive production of $A$ 
were computed and it was found that the scale dependence reduced from 48\% to 35\% and
the corrections were found to be as large as 67\%.  The effective field theory approach
where the top quark degrees of freedom are integrated out has provided opportunity to
go beyond NLO level to further improve the predictions. 
The results on NNLO production cross section for $A$ can be found
in ~\cite{Harlander:2002vv,Anastasiou:2002wq,Ravindran:2003um} which further improves the
reliability of the predictions.  

There have been continued efforts to go beyond NNLO
as the results on the form factors (FFs) and soft gluon contributions at third order level
have become available.  For this purpose, in \cite{Ahmed:2015qpa}, we computed   
FFs of the effective composite operators  
between quark and gluon states at three loop level in QCD along with the lower order.
Using 
the formalism developed in ~\cite{Ravindran:2005vv,
  Ravindran:2006cg} and the third order results on  
the FFs of the pseudo-scalar Higgs
boson~\cite{Ahmed:2015qpa}, the universal soft-collinear
distribution~\cite{Ahmed:2014cla}, operator renormalisation
constant~\cite{Larin:1993tq, Zoller:2013ixa, Ahmed:2015qpa} and the
mass factorisation kernels~\cite{Vogt:2004mw, Moch:2004pa}, the first results 
on the threshold N${}^3$LO corrections for the inclusive production of $A$ were reported in 
\cite{Ahmed:2015qda}.  In addition, 
the third order corrections to both
the $N$-independent part of the resummed cross
section~\cite{Sterman:1986aj,Catani:1989ne} and the matching coefficients in soft-collinear effective theory (SCET) 
were also computed in \cite{Ahmed:2015qpa}.  In \cite{Ahmed:2016otz},  
we obtained the approximate N${}^3$LO contribution using the full N${}^3$LO results available for the 
scalar Higgs boson.  This along with the 
threshold effects at next-to-next-to-next-to-leading logarithm (N${}^3$LL) accuracy
using both conventional ~\cite{Sterman:1986aj,
Catani:1989ne} and SCET ~\cite{Bauer:2000ew, Bauer:2000yr, Bauer:2001ct,Bauer:2001yt,Beneke:2002ph,
Beneke:2002ni,Bauer:2002nz} setups provide the accurate prediction \cite{Ahmed:2016otz} 
for the inclusive production of $A$ at the LHC. 

In order to probe the nature of $A$ and its coupling to other SM particles,
one also needs to study exclusive observables namely the differential cross sections.  
The distributions of transverse momentum and rapidity of the produced $A$ 
are often very useful for characterising its properties.  
In addition, observables involving $A$ recoiled against one or two hard jets
can help to efficiently reject the background from other sources.  
There are already several studies in the case of scalar Higgs boson in the literature. 
For example, the differential cross sections for scalar Higgs boson 
via gluon fusion are available including its decay to two photons 
or two weak gauge bosons, see 
~\cite{Anastasiou:2005qj,Anastasiou:2007mz,Grazzini:2008tf}.  
It was noted in ~\cite{Catani:2001cr,Berger:2010xi} 
that observables with jet vetos  enhance the significance of the signal 
considerably allowing one to study the 
properties of the Higgs boson and its coupling to other SM particles. 
Recently, in \cite{Chen:2016zka}, a complete NNLO prediction for Higgs-plus-jet final states and 
for the transverse momentum distribution of the Higgs boson taking into account 
the experimental definition of the fiducial cross sections has been achieved. 
For Higgs-plus-jet production at NNLO, see
\cite{Boughezal:2013uia,Chen:2014gva,Boughezal:2015dra,Boughezal:2015aha,Caola:2015wna}.
The relevant two loop amplitudes can be found in 
~\cite{Dixon:2004za,Badger:2004ty,Badger:2006us,
Dixon:2009uk,Badger:2009hw,Badger:2009vh,Gehrmann:2011aa}. 
Pseudo-scalar Higgs boson being produced dominantly in gluon initiated processes
shares a lot with its counter part, namely the scalar Higgs boson.   
In particular one finds that the theoretical uncertainties 
and the perturbative corrections are significantly large.  
For the pseudo-scalar Higgs boson, the results for differential distributions 
are available only up to NLO level, for example, see \cite{Field:2002pb,Bernreuther:2010uw}. 
Hence it is desirable to have predictions upto NNLO level for observables like $A +1$jet,
transverse momentum and rapidity distributions of $A$, taking into account various experimental
cuts.  At NNLO, one encounters three different production channels that contribute to these
observables: pure virtual, virtual-real emissions, pure real emissions.  
In this paper, we will present 
all the relevant contributions resulting from pure virtual amplitudes. 
In particular, we will consider the 
processes $A\rightarrow ggg$ and $A\rightarrow q\bar{q}g $ in effective theory where
top quark has been integrated out up to two loop level. 
These amplitudes can be used to predict the production of $A$ in hadron collider 
up to two loop level in QCD after performing straightforward crossing of kinematic variables.
In the effective theory for pseudo-scalar, 
one is left with two effective operators with same quantum numbers and mass 
dimensions, hence they mix under renormalisation. 
We will present the results in the ${\overline {MS}}$ scheme. 

Our paper is organised as follows.  In Section \ref{sec:efflag}, the effective Lagrangian for the coupling of $A$ with
partons along with the relevant Wilson coefficients are presented.  
After introducing the notation in Section \ref{sec:nota}, we elaborate on how to deal with $\gamma_5$ and Levi-Civita tensor
in dimensional regularisation and subtleties involved in operator mixing and finite renormalisation in Section \ref{sec:reamp}. 
In Section \ref{sec:mtelements}, we describe the method of our computation of relevant matrix elements up to two loop level in QCD and Section \ref{sec:usirv}
contains the study of IR structure of the results obtained in the context of Catani's predictions.  In Section \ref{sec:LT},
we discuss our results on one and two loops QCD amplitudes in the light of maximum transcendentality principle in
${\cal N}=4$ SYM.  Finally we conclude in Section \ref{sec:con}.
 
 %****Theory****
\section{Theoretical Framework}
\label{sec:theory}
\subsection{Effective Lagrangian}
\label{sec:efflag}
In the generic BSM scenarios, the pseudo-scalar Higgs boson couples to heavy quarks through Yukawa interaction. 
If we restrict ourselves to the dominant contribution coming from top Yukawa coupling, then the interaction Lagrangian   
is given by
\begin{eqnarray}
{\cal L}_{At\overline t} = - i g_t {A\over v} m_t \overline t \gamma_5 t
\end{eqnarray}  
where the model dependent constant $g_t$ in MSSM is $\cos\beta$ where $\tan \beta$ is ratio of vacuum expectation values of
the two Higgs fields, $m_t$ is the top quark mass, $v$ the SM vacuum expectation value. 

In the limit of infinite top quark mass, the interaction between a pseudo-scalar 
$A$ and the fields of the remaining SM is encapsulated by an effective 
Lagrangian~\cite{Chetyrkin:1998mw} which reads as
\begin{align} {\cal L}^{A}_{\rm eff} = \Phi^{A}(x) \Big[ -\frac{1}{8}
  {C}_{G} O_{G}(x) - \frac{1}{2} {C}_{J} O_{J}(x)\Big]
  \label{eq:efflag}
\end{align}
where the operators are defined as
\begin{equation}
  O_{G}(x) = G^{\mu\nu}_a \tilde{G}_{a,\mu
    \nu} \equiv  \epsilon_{\mu \nu \rho \sigma} G^{\mu\nu}_a G^{\rho
    \sigma}_a\, ,\qquad
  O_{J}(x) = \partial_{\mu} \left( \bar{\psi}
    \gamma^{\mu}\gamma_5 \psi \right)  \,.
  \label{eq:operators}
\end{equation}
The Wilson coefficients $C_G$ and $C_J$ are obtained by integrating
out the loops resulting from top quark. The coefficient $C_G$ does not receive any QCD
corrections beyond one loop in the perturbation series expanded in strong coupling constant $a_s \equiv 
{g}_{s}^{2}/(16\pi^{2}) = \alpha_s/(4\pi)$ due to the Adler-Bardeen
theorem~\cite{Adler:1969gk} and $C_J$ starts at second order. Their 
expressions are given below
\begin{align}
  \label{eq:const}
  & C_{G} = -a_{s} 2^{\frac{5}{4}} G_{F}^{\frac{1}{2}} {\rm \cot} \beta,
    \nonumber\\
  & C_{J} = - \left[ a_{s} C_{F} \left( \frac{3}{2} - 3\ln
    \frac{\mu_{R}^{2}}{m_{t}^{2}} \right) + a_s^2 C_J^{(2)} + \cdots \right] 
C_{G}\, .
\end{align}
In the above expressions, $G^{\mu\nu}_{a}$ and $\psi$ represent
gluonic field strength tensor and light quark fields, respectively and
$G_{F}$ is the Fermi constant. Here,
$a_{s} \equiv a_{s} \left( \mu_{R}^{2} \right)$ is the renormalised strong 
coupling
constant at the scale $\mu_{R}$ which is related to the
unrenormalised one, ${\hat a}_{s} \equiv {\hat g}_{s}^{2}/(16\pi^{2})$
through
\begin{align}
  \label{eq:asAasc}
  {\hat a}_{s} S_{\epsilon} = \left( \frac{\mu^{2}}{\mu_{R}^{2}}  
\right)^{\epsilon/2}
  Z_{a_{s}} a_{s}
\end{align}
with
$S_{\epsilon} = {\rm exp} \left[ (\gamma_{E} - \ln 4\pi)\epsilon/2
\right]$
and $\mu$ is the scale introduced to keep the strong coupling constant
dimensionless in $d=4+\epsilon$ space-time dimensions.  The
renormalisation constant $Z_{a_{s}}$~\cite{Tarasov:1980au} up to order ${\cal 
O}(a_{s}^{2})$ is given by
\begin{align}
  \label{eq:Zas}
  Z_{a_{s}}&= 1+ a_s\left[\frac{2}{\epsilon} \beta_0\right]
             + a_s^2 \left[\frac{4}{\epsilon^2 } \beta_0^2
             + \frac{1}{\epsilon}  \beta_1 \right]
\end{align}
 $\beta_{i}$ are the coefficients of the
QCD $\beta$-functions~\cite{Tarasov:1980au}, given by
\begin{align}
  \beta_0&={11 \over 3 } C_A - {4 \over 3 } n_f T_{F}\, ,
           \nonumber \\[0.5ex]
  \beta_1&={34 \over 3 } C_A^2- 4 n_f C_F T_{F} -{20 \over 3} n_f
           T_{F} C_A \, ,
\end{align}
with the SU(N) QCD color factors
\begin{equation}
  C_A=N,\quad \quad C_F={N^2-1 \over 2 N} \quad \text{and} \quad T_{F}= 
\frac{1}{2}\,.
\end{equation}
$n_f$ is the number of active light quark flavours.
\subsection{Notation}
\label{sec:nota}
We consider the decay of pseudo-scalar Higgs boson to both $ggg$ and 
$q\bar{q}g$ which can be expressed as
\begin{align}
A(q) &\rightarrow g(p_1) + g(p_2) + g(p_3),
\nonumber\\
A(q) &\rightarrow q(p_1) + \bar{q}(p_2) + g(p_3).
\end{align}
The associated Mandelstam variables are defined as
\begin{equation}
s \equiv (p_1+p_2)^2 > 0,\qquad  t \equiv (p_2+p_3)^2 >0,\qquad u \equiv 
(p_1+p_3)^2 > 0,
\end{equation}
which satisfy the relation 

\begin{equation}
s + t + u = M_A^2 \equiv q^2 = Q^2 > 0,
\end{equation}
where $M_A$ is the mass of the pseudo-scalar Higgs boson. We also define 
following dimensionless invariants, which we use to describe our results in 
terms of HPLs~\cite{Remiddi:1999ew} and 
2d-HPLs~\cite{Gehrmann:2000zt,Gehrmann:2001ck} as

\begin{equation}
x \equiv \frac{s}{Q^2}, \qquad y \equiv \frac{u}{Q^2},  \qquad  z \equiv 
\frac{t}{Q^2}, 
\end{equation}

where $(x,y,z)$ lie between 0 and 1 and satisfy the condition 
\begin{equation}
x+ y + z=1\,.
\end{equation}
\subsection{Operator mixing and UV renormalisation}
\label{sec:reamp}
In the following, we will describe the computation of one and two loop matrix elements
that are needed to perform NNLO QCD corrections to
production of a pseudo-scalar with a jet in hadron colliders or its decay to three jets.  
Since the amplitudes for the production of pseudo-scalar in association with a jet
 from gluon gluon or quark anti-quark initiated channels are related to
the decay of the same to three gluons or quark anti-quark gluon by crossing symmetry,
in the rest of the paper we will only describe the latter in detail.
The composite operators present in the effective Lagrangian Eq.~(\ref{eq:efflag}),  
develop UV divergences that require additional renormalisation. Also these operators mix under renormalisation due to same quantum numbers.

In higher order computations involving chiral quantities,
the inherently four dimensional objects like  $\gamma_5$ and 
$\varepsilon^{\mu\nu\rho\sigma}$, the Levi-Civita tensor, 
in $d\neq 4$ dimensions pose problems. 
In this article, we have followed the most practical and
self-consistent definition of $\gamma_{5}$ introduced for multi loop calculations
in dimensional regularization by 't~Hooft and
Veltman through \cite{tHooft:1972tcz}
\begin{align}
  \gamma_5 = i \frac{1}{4!} \varepsilon_{\nu_1 \nu_2 \nu_3 \nu_4}
  \gamma^{\nu_1}  \gamma^{\nu_2} \gamma^{\nu_3} \gamma^{\nu_4} \,.
\end{align}
Here, $\varepsilon^{\mu\nu\rho\sigma}$ is 
contracted according to the rule 
\begin{align}
  \label{eqn:LeviContract}
\varepsilon_{\mu_1\nu_1\rho_1\sigma_1}\,\varepsilon^{\mu_2\nu_2\rho_2\sigma_2}=
  \large{\left |
  \begin{array}{cccc}
    \delta_{\mu_1}^{\mu_2} &\delta_{\mu_1}^{\nu_2}&\delta_{\mu_1}^{\rho_2} & 
\delta_{\mu_1}^{\sigma_2}\\
\delta_{\nu_1}^{\mu_2}&\delta_{\nu_1}^{\nu_2}&\delta_{\nu_1}^{\rho_2}&\delta_{\nu_1}^{\sigma_2}\\
\delta_{\rho_1}^{\mu_2}&\delta_{\rho_1}^{\nu_2}&\delta_{\rho_1}^{\rho_2}&\delta_{\rho_1}^{\sigma_2}\\
\delta_{\sigma_1}^{\mu_2}&\delta_{\sigma_1}^{\nu_2}&\delta_{\sigma_1}^{\rho_2}&
\delta_{\sigma_1}^{\sigma_2}
  \end{array}
       \right |}
\end{align}
where all the Lorentz indices are considered to be 
$d$-dimensional~\cite{Larin:1993tq}. 
The prescription used here fails to preserve the anti-commutativity of $\gamma_{5}$ with 
$\gamma^{\mu}$ in arbitrary $d$-dimensions.  In addition, 
the Ward identities, which are valid in a 4-dimensional regularization 
scheme like the one of
Pauli-Villars where $\gamma_5$ does not pose any problem, are violated as well. 
Due to this, it is not possible to restore the 
correct renormalisation of axial current, which is defined as \cite{Larin:1993tq, 
Akyeampong:1973xi}.
A finite renormalisation of the 
axial vector current is required to preserve chiral Ward identities and 
the Adler-Bardeen theorem.  The axial current is defined as
\begin{align}
  \label{eq:J5}
  J^{\mu}_{5} \equiv \bar{\psi}\gamma^{\mu}\gamma_{5}\psi = i
  \frac{1}{3!} \varepsilon^{\mu\nu_{1}\nu_{2}\nu_{3}} \bar{\psi}
  \gamma_{\nu_{1}} \gamma_{\nu_{2}}\gamma_{\nu_{3}} \psi
\end{align}
in dimensional regularization.  The chiral Ward identities can be 
fixed by introducing a finite 
renormalisation constant $Z^{s}_{5}$
\cite{Adler:1969gk,Kodaira:1979pa} in addition to the standard overall UV 
renormalisation constant $Z^{s}_{\overline{MS}}$ within the 
$\overline{MS}$-scheme:
\begin{align}
  \label{eq:J5Ren}
  \left[ J^{\mu}_{5} \right]_{R} = Z^{s}_{5} Z^{s}_{\overline{MS}} \left[ 
J^{\mu}_{5} \right]_{B}\,.
\end{align}
While the evaluation of the appropriate Feynman diagrams explicitly fixes
the constants $Z^{s}_{\overline{MS}}$, the finite
renormalisation constant can not be fixed without using the anomaly equation. In other words, 
$Z^{s}_{5}$ can be determined by demanding the conservation of the one loop character 
\cite{Adler:1969er} of the operator relation of the axial anomaly in 
dimensional regularization:
\begin{align}
  \label{eq:Anomaly}
  \left[ \partial_{\mu}J^{\mu}_{5} \right]_{R} &= a_{s} \frac{n_{f}}{2} \left[ 
G\tilde{G} \right]_{R}
  \nonumber\\
  \text{i.e.}~~~ \left[ O_{J} \right]_{R} &= a_{s} \frac{n_{f}}{2} \left[ O_{G} 
\right]_{R}\,.
\end{align}
The bare operator $\left[ O_{J} \right]_{B}$ is renormalised multiplicatively 
as the axial current $J^{\mu}_{5}$ through
\begin{align}
  \label{eq:OJRen}
  \left[ O_{J} \right]_{R} = Z^{s}_{5} Z^{s}_{\overline{MS}} \left[ 
O_{J}\right]_{B}\,,
\end{align}
but the other bare operator $\left[ O_{G} \right]_{B}$ mixes under the
renormalisation through
\begin{align}
  \left[ O_{G} \right]_{R} = Z_{GG} \left[ O_{G}\right]_B +
  Z_{GJ} \left[ O_{J} \right]_B
\end{align}
through renormalisation constants $Z_{GG}$ and
$Z_{GJ}$. The above two equations can be written as
\begin{align}
  \label{eq:OpMat}
  \left[ O_{\Sigma} \right]_{R} &= Z_{\Sigma \Lambda} \left[  O_{\Lambda}\right]_{B} 
\end{align}
with
\begin{align}
  \label{eq:ZMat}
  \Sigma, \Lambda &= \{G, J\}\,, 
        \nonumber\\
  O \equiv
  \begin{bmatrix}
    O_{G}\\
    O_{J}
  \end{bmatrix}
  \qquad\quad &\text{and}  \qquad\quad
                Z \equiv
                \begin{bmatrix}
                  Z_{GG} & Z_{GJ}\\
                  Z_{JG} & Z_{JJ}
                \end{bmatrix}\,,
\end{align}
where
\begin{align}
  \label{eq:ZJGZJJ}
  Z_{JG} &= 0 \qquad \text{to all orders in perturbation theory}\,,
           \nonumber\\
  Z_{JJ} &\equiv Z^{s}_{5} Z^{s}_{\overline{MS}}\,.
\end{align}
The expressions for the above mentioned renormalisation constants 
$Z^{s}_{\overline{MS}},Z_{GG},Z_{GJ}$ up to
${\cal{O}}\left( a_{s}^{3} \right)$ are given 
in~\cite{Larin:1993tq,Zoller:2013ixa} using operator product expansion. 
In~\cite{Ahmed:2015qpa} one of us has calculated the 
same quantities in a completely different way and found exact agreement with the 
original calculation. In the latter article, authors have used universality of 
the IR poles of the FF to determine the UV renormalisation constants and 
also computed $Z_5^s$ up to $\mathcal{O}(a_s^2)$ by demanding the operator relation of the axial anomaly 
Eq.~(\ref{eq:Anomaly}). These renormalisation constants up to sufficient order 
in the perturbation theory appropriate for our calculation are given below:

\begin{align}
  \label{eq:ZGGtZGJ}
  Z_{GG} &= 1 +  a_s \Bigg[ \frac{22}{3\epsilon}
           C_{A}  -
           \frac{4}{3\epsilon} n_{f} \Bigg] 
           + 
           a_s^2 \Bigg[ \frac{1}{\epsilon^2}
           \Bigg\{ \frac{484}{9} C_{A}^2 - \frac{176}{9} C_{A}
           n_{f} + \frac{16}{9} n_{f}^2 \Bigg\}
           \nonumber\\&
           + \frac{1}{\epsilon} \Bigg\{ \frac{34}{3} C_{A}^2  
         -\frac{10}{3} C_{A} n_{f}  - 2 C_{F} n_{f} \Bigg\} \Bigg],
         \nonumber\\
  Z_{GJ} &=  a_s \Bigg[ - \frac{24}{\epsilon} C_{F} \Bigg]
                    + 
                    a_s^2 \Bigg[ \frac{1}{\epsilon^2}
                    \Bigg\{ - 176 C_{A} C_{F} + 32 C_{F} n_{f} \Bigg\}
                    \nonumber\\&
                    + \frac{1}{\epsilon} \Bigg\{ - \frac{284}{3} C_{A} C_{F} +
                    84 C_{F}^2 + \frac{8}{3} C_{F} n_{f} \Bigg\}  \Bigg],
                    \nonumber\\
   Z_{JJ} &= 1 + a_s\left[-4 C_F\right] + a_s^2\Bigg[  
-\frac{44}{3\epsilon}C_A C_F - \frac{10}{3\epsilon}C_F n_f   
   \nonumber\\&
   + 22C_F^2 -\frac{107}{9}C_A C_F  + \frac{31}{18}C_F n_f \Bigg].
\end{align}
Using these operator renormalisation constants along with strong coupling 
constant renormalisation through $Z_{a_s}$, we obtain UV finite amplitudes. 

\subsection{Matrix elements}
\label{sec:mtelements}
Our next step is to compute all the relevant matrix elements resulting from virtual amplitudes
for the decay of pseudo-scalar to three gluons and also to quark antiquark gluon.
They are obtained from the amplitudes $|{\cal A}_{f}\rangle$, where $f=ggg,q \bar q g$
up to two loop level.  They contain two sub-amplitudes namely $|\mathcal{M}^{\Lambda}_f\rangle $ 
computed using $O_G$ ($\Lambda =G$)  and $O_J$ ($\Lambda =J$) operators
multiplied by the appropriate Wilson coefficients $C_\Lambda$: 
\begin{eqnarray}
|\mathcal{A}_f\rangle=\sum_{\Lambda=G,J}  C_\Lambda(a_s) |\mathcal{M}^{\Lambda}_f\rangle   
\end{eqnarray}
The above UV finite sub-amplitudes $|\mathcal{M}^{\Lambda}_f\rangle$ 
can be expressed in terms unrenormalised sub-amplitudes $|\hat {\mathcal{M}}^{\Lambda,(n)}_f\rangle$ as follows:
\begin{equation}
|\mathcal{M}^{\Lambda}_f\rangle = (16 \pi^2 a_s)^{\frac{1}{2}}\sum_{\Sigma=G,J} Z_{\Lambda \Sigma}\left( 
|\mathcal{M}^{\Sigma,(0)}_f\rangle + a_s|\mathcal{M}^{\Sigma,(1)}_f\rangle + 
a_s^2|\mathcal{M}^{\Sigma,(2)}_f\rangle + \mathcal{O}(a_s^3)\right),
\end{equation}
where
\begin{align}
|\mathcal{M}^{\Sigma,(0)}_f\rangle =& \left( 
\frac{1}{\mu_R^{\epsilon}}\right)^{\frac{1}{2}}|\hat{\mathcal{M}}^{\Sigma,(0)}_f
\rangle,
\nonumber\\
|\mathcal{M}^{\Sigma,(1)}_f\rangle =&\left( 
\frac{1}{\mu_R^{\epsilon}}\right)^{\frac{3}{2}}\left[ 
|\hat{\mathcal{M}}^{\Sigma,(1)}_f\rangle 
+\mu_R^{\epsilon}\frac{r_1}{2}|\hat{\mathcal{M}}^{\Sigma,(0)}_f\rangle \right],
\nonumber\\
|\mathcal{M}^{\Sigma,(2)}_f\rangle =&\left( 
\frac{1}{\mu_R^{\epsilon}}\right)^{\frac{5}{2}}\left[ 
|\hat{\mathcal{M}}^{\Sigma,(2)}_f\rangle  + 
\mu_R^{\epsilon}\frac{3r_1}{2}|\hat{\mathcal{M}}^{\Sigma,(1)}_f\rangle + 
\mu_R^{2\epsilon}\left( \frac{r_2}{2}-\frac{r_1^2}{8}\right) 
|\hat{\mathcal{M}}^{\Sigma,(0)}_f\rangle\right]
\end{align}
with
\begin{equation}
r_1 = \frac{2\beta_0}{\epsilon}, \qquad\qquad r_2 = \left(  
\frac{4\beta_0^2}{\epsilon^2} + \frac{\beta_1}{\epsilon} \right)\,.
\end{equation}
In the above equations, the unrenormalised sub-amplitudes $|\hat {\cal M}_f^\Sigma \rangle$ are computed 
in powers of bare coupling constant $\hat a_s$. 
Using these UV finite sub-amplitudes $|{\cal M}_f^\Sigma \rangle$, we then obtain relevant matrix element squares to
observables to desired order in $a_s$.  
For $A \rightarrow ggg$, we find
\begin{eqnarray}
{\cal S}_{ggg} &=&  \langle\mathcal{A}_{ggg}|\mathcal{A}_{ggg}\rangle 
\nonumber \\
&=& a_s^3 \left(C_G^{(1)}\right)^2 \Bigg[{S}_g^{G,(0)} + a_s {S}_g^{G,(1)}
+ a_s^2 \left( {S}_g^{G,(2)} + 2 C_J^{(1)} {S}_g^{GJ,(1)} \right) \Bigg] 
\end{eqnarray}
where
\begin{eqnarray}
{S}_g^{G,(0)} &=& \langle \mathcal{M}^{G,(0)}_{ggg} | \mathcal{M}^{G,(0)}_{ggg}\rangle \, , 
\nonumber\\
{S}_g^{G,(1)} &=&  2 \langle \mathcal{M}^{G,(0)}_{ggg} | \mathcal{M}^{G,(1)}_{ggg}\rangle \, ,
\nonumber\\
{S}_g^{GJ,(1)} &=& \langle \mathcal{M}^{G,(0)}_{ggg} | \mathcal{M}^{J,(1)}_{ggg}\rangle  \, ,
\nonumber\\
{S}_g^{G,(2)} &=&
\langle \mathcal{M}^{G,(1)}_{ggg} | \mathcal{M}^{G,(1)}_{ggg}\rangle
+2 \langle \mathcal{M}^{G,(0)}_{ggg} | \mathcal{M}^{G,(2)}_{ggg}\rangle 
\end{eqnarray}
and similarly for $A \rightarrow q \overline q g$,  we find
\begin{eqnarray} 
{\cal S}_{q\bar{q}g} &=& \langle\mathcal{A}_{q\bar{q}g}|\mathcal{A}_{q\bar{q}g}\rangle
\nonumber\\
&=& a_s^3 \left(C_G^{(1)}\right)^2 \Bigg[ {S}_q^{G,(0)} 
   + a_s \left(2 {S}_q^{G,(1)} + 2 C_J^{(1)} {S}_q^{JG,(0)}\right)
\nonumber\\
&&   + a_s^2 \left( {S}_q^{G,(2)} 
          + C_J^{(1)} {S}_q^{JG,(1)} 
          + \left(C_J^{(1)}\right)^2 {S}_q^{J,(0)} + 2 C_J^{(2)} {S}_q^{JG,(0)}\right) \Bigg]
\end{eqnarray}
where
\begin{eqnarray}
{S}_q^{G,(0)} &=& \langle \mathcal{M}^{G,(0)}_{q\bar{q}g} | \mathcal{M}^{G,(0)}_{q\bar{q}g}\rangle \, ,
\nonumber\\
{S}_q^{J,(0)} &=& \langle \mathcal{M}^{J,(0)}_{q\bar{q}g} | \mathcal{M}^{J,(0)}_{q\bar{q}g}\rangle  \, ,
\nonumber\\
{S}_q^{G,(1)} &=&  \langle \mathcal{M}^{G,(0)}_{q\bar{q}g} | \mathcal{M}^{G,(1)}_{q\bar{q}g}\rangle \, ,
\nonumber\\
{S}_q^{JG,(0)}&=& \langle \mathcal{M}^{J,(0)}_{q\bar{q}g} | \mathcal{M}^{G,(0)}_{q\bar{q}g} \rangle \, ,
\nonumber\\
{S}_q^{G,(2)} &=&\langle \mathcal{M}^{G,(1)}_{q\bar{q}g} | \mathcal{M}^{G,(1)}_{q\bar{q}g} \rangle 
          + 2 \langle \mathcal{M}^{G,(0)}_{q\bar{q}g} | \mathcal{M}^{G,(2)}_{q\bar{q}g} \rangle  \, ,
\nonumber\\
{S}_q^{JG,(1)}&=& 2\langle  \mathcal{M}^{J,(1)}_{q\bar{q}g} | \mathcal{M}^{G,(0)}_{q\bar{q}g} \rangle  
+ 2\langle \mathcal{M}^{J,(0)}_{q\bar{q}g} | \mathcal{M}^{G,(1)}_{q\bar{q}g} \rangle \, ,
\end{eqnarray}
and
\begin{align}
  \label{eq:const}
   C_{G}^{(1)} = - 2^{\frac{5}{4}} G_{F}^{\frac{1}{2}} {\rm \cot} \beta \,,
\quad \quad
   C_{J}^{(1)} = - C_{F} \left( \frac{3}{2} - 3\ln
    \frac{\mu_{R}^{2}}{m_{t}^{2}} \right)  .
\end{align}
The computation of $S_a^{\Lambda,(i)}$ for $a = g,q;\Lambda=G,J,GJ,JG,i=0,1,2$ closely follows the steps used in
~\cite{deFlorian:2013sza,deFlorian:2013wpa,Gehrmann:2014vha,Ahmed:2014gla,Ahmed:2014pka,Ahmed:2015qia,Ahmed:2016qhu,Ahmed:2016yox}. We briefly describe the main steps involved.   
The relevant tree, one and two loop Feynman
diagrams are generated using the package QGRAF~\cite{Nogueira:1991ex}.  While there are only few diagrams at tree and one loop level,
at two loops we encounter large number of diagrams.  We find that the number of two loop diagrams  
is 1306 for $| 
\hat{\cal{M}}_{ggg}^{G,(2)}\rangle$, 264 for $| 
\hat{\cal{M}}_{ggg}^{J,(2)}\rangle$, 328 for $| 
\hat{\cal{M}}_{q\bar{q}g}^{G,(2)}\rangle$ and 229 for $| 
\hat{\cal{M}}_{q\bar{q}g}^{J,(2)}\rangle$.  We set all the external particles 
on-shell, in other words, the quarks, anti-quarks, gluons are kept massless and the pseudo-scalar has non-zero mass $M_A$. 
The raw QGRAF output is converted to a
format that can be further used to perform the substitution of Feynman rules,
contraction of Lorentz and color indices and simplification of Dirac and
Gell-Mann matrices using a set of in-house routines written in
the symbolic manipulating program FORM \cite{Vermaseren:2000nd}.  We have
included ghost loops in the Feynman gauge.  For the external on-shell gluons,
we have kept only transversely polarization states in
d-dimensions. The next step  is to organise all the d-dimensional one and two loop integrals
such that they can be reduced to a
minimum set of scalar integrals, called master integrals (MIs).  To do this,
 we first use Reduze2~\cite{vonManteuffel:2012np, Studerus:2009ye} to shift loop momenta to
get suitable integral classes.   
Each integral belonging to specific class can be expressed in terms of MIs using a set of 
integration-by-parts (IBP) \cite{Tkachov:1981wb,Chetyrkin:1981qh} and Lorentz
invariance (LI) \cite{Gehrmann:1999as} identities.  The LI identities are
not linearly independent from the IBP identities~\cite{Lee:2008tj}, 
however they help to increase the speed to solve the large number of linear equations resulting
from IBP. The method of IBP to reduce certain class of integrals to a set of MIs 
is achieved using Laporta
algorithm,~\cite{Laporta:2001dd}.  This has been implemented in various symbolic manipulation
packages such as  AIR~\cite{Anastasiou:2004vj}, FIRE~\cite{Smirnov:2008iw},
Reduze2~\cite{vonManteuffel:2012np, Studerus:2009ye} and
LiteRed~\cite{Lee:2013mka, Lee:2012cn}.  
We have used 
LiteRed~\cite{Lee:2013mka, Lee:2012cn} to perform the reductions of all the
integrals to MIs.  For one and two loop, we find the number of MIs are 7 and 89
respectively.  The next task is to express all the MIs that we obtain 
after using LiteRed to those computed analytically in
~\cite{Gehrmann:2000zt,Gehrmann:2001ck}.   Using these MIs in the appropriate kinematic
regions, we obtain analytical results for ${\cal S}_f$ for $f=ggg,q\bar q g$.
While these amplitude squares are UV finite, they are sensitive to IR divergences.
Thanks to universality of these divergences, our results can be verified up to finite terms.

\subsection{Universal Structure of IR divergences}
\label{sec:usirv}
Beyond leading order in perturbation theory, the UV renormalised amplitudes in gauge theories suffer from IR 
divergences due to the presence of massless particles. These 
divergences are classified into two categories, namely soft and collinear. Soft divergences arise 
when momentum of a massless particle in the loop goes to zero and the collinear 
ones result when a massless particle in the loop becomes parallel to one of the 
external particles. Note that the amplitudes are not observables, instead 
the cross sections or decay rates made out of these amplitudes are the 
observables. Thanks to Kinoshita-Lee-Nauenberg (KLN) theorem 
~\cite{Kinoshita:1962ur,Lee:1964is}, the finiteness of the 
observables can be guaranteed by summing over the degenerate final states and performing mass factorisation for the 
initial states. While these divergences go away in the physical observables,
the amplitudes demonstrate very rich universal structure in the IR region at every order in
perturbation theory.  The IR structure of 
amplitudes in QCD is well understood and in particular, the universal structure 
of IR poles was predicted in a seminal paper ~\cite{Catani:1998bh} by Catani 
for n-point two loop amplitudes. The iterative structure of 
singular part of the UV renormalised amplitudes in QCD was exploited in ~\cite{Catani:1998bh} to predict the 
subtraction operators that capture the universal IR divergences. Later on 
Sterman and Tejeda-Yeomans related the predictions by Catani to the 
factorisation and resummation properties of QCD amplitudes 
~\cite{Sterman:2002qn}.  It is easy to convince oneself that iterative structure 
is the result of factorisation.  
The generalization of Catani's proposal for arbitrary number of loops and legs 
for $SU(N)$ gauge theory
 using soft-collinear effective theory was given by Becher and Neubert 
~\cite{Becher:2009cu}.  A similar result was also presented by Gardi and Magnea 
~\cite{Gardi:2009qi} by analyzing Wilson lines for hard partons.
 
 We follow here Catani's proposal and express the UV renormalised amplitudes 
$| \mathcal{M}^{\Sigma,(n)}_f \rangle$ for the three point function up to 
two loop order in terms of universal subtraction operators 
$\mathbf{I}^{(n)}_f(\epsilon)$,
 \begin{align}
 | \mathcal{M}^{\Sigma,(1)}_f \rangle &=2 \mathbf{I}^{(1)}_f(\epsilon)|\mathcal{M}^{\Sigma,(0)}_f\rangle + 
|\mathcal{M}^{\Sigma,(1)fin}_f \rangle,
 \nonumber\\
 | \mathcal{M}^{\Sigma,(2)}_f \rangle &=4\mathbf{I}^{(2)}_f(\epsilon) |\mathcal{M}^{\Sigma,(0)}_f \rangle  + 
2\mathbf{I}^{(1)}_f(\epsilon)| \mathcal{M}^{\Sigma,(1)}_f \rangle + | \mathcal{M}^{\Sigma,(2)fin}_f \rangle\,,
 \end{align}
 where 
 \begin{align}
 \label{Reg9}
 \mathbf{I}_{ggg}^{(1)}(\epsilon) =& 
-\frac{1}{2}\frac{e^{-\frac{\epsilon}{2}\gamma_E}}{\Gamma(1+\frac{\epsilon}{2})}
 \left( C_A\frac{4}{\epsilon^2}-\frac{\beta_0}{\epsilon} 
\right)\left[\Big(-\frac{s}{\mu_R^2}\Big)^{\frac{\epsilon}{2}} + 
\Big(-\frac{t}{\mu_R^2}\Big)^{\frac{\epsilon}{2}} +
  \Big(-\frac{u}{\mu_R^2}\Big)^{\frac{\epsilon}{2}}\right] \, ,
  \nonumber\\
 \mathbf{I}_{q\bar{q}g}^{(1)}(\epsilon) =& 
-\frac{1}{2}\frac{e^{-\frac{\epsilon}{2}\gamma_E}}{\Gamma(1+\frac{\epsilon}{2})}
\Bigg\{ 
\left(\frac{4}{\epsilon^2}-\frac{3}{\epsilon}\right)(C_A-2C_F)\left[\left(-\frac
{s}{\mu_R^2}\right)^{\frac{\epsilon}{2}}\right]  
 \nonumber\\&
+\left(-\frac{4C_A}{\epsilon^2}+\frac{3C_A}{2\epsilon}+\frac{\beta_0}{2\epsilon}
\right)\left[\left(-\frac{t}{\mu_R^2}\right)^{\frac{\epsilon}{2}} + 
\left(-\frac{u}{\mu_R^2}\right)^{\frac{\epsilon}{2}}\right]  \Bigg\}\,,
 \nonumber\\
 \mathbf{I}_{f}^{(2)}(\epsilon) =& 
-\frac{1}{2}\mathbf{I}^{(1)}_f(\epsilon)\left[ \mathbf{I}^{(1)}_f(\epsilon) - 
\frac{2\beta_0}{\epsilon} \right] + 
\frac{e^{\frac{\epsilon}{2}\gamma_E}\Gamma(1+\epsilon)}{\Gamma(1+\frac{\epsilon}
{2})}\left[-\frac{\beta_0}{\epsilon}+K\right]\mathbf{I}_{f}^{(1)}(2\epsilon)+
\mathbf{H}^{(2)}_f(\epsilon)
 \end{align}
 with
 \begin{align}
 K =&\left( \frac{67}{18}-\zeta_2\right)C_A - \frac{10}{9}T_F n_f,
 \nonumber\\
 \mathbf{H}^{(2)}_{ggg}(\epsilon) =&\frac{3}{\epsilon}\left\{ 
C_A^2\left(-\frac{5}{24}-\frac{11}{48}\zeta_2-\frac{1}{4}\zeta_3\right) + C_A 
n_f\left(\frac{29}{54} + \frac{1}{24}\zeta_2\right)-\frac{1}{4}C_F n_f 
-\frac{5}{54}n_f^2\right\}\,,
 \nonumber\\
 \mathbf{H}^{(2)}_{q\bar{q}g}(\epsilon) =&\frac{1}{\epsilon}\Bigg\{ 
C_A^2\left(-\frac{5}{24}-\frac{11}{48}\zeta_2-\frac{1}{4}\zeta_3\right) + C_A 
n_f\left(\frac{29}{54}+\frac{1}{24}\zeta_2\right)+C_F 
n_f\left(-\frac{1}{54}-\frac{1}{4}\zeta_2\right)    
 \nonumber\\&
 +C_A C_F\left(-\frac{245}{216} + 
\frac{23}{8}\zeta_2-\frac{13}{2}\zeta_3\right) + 
C_F^2\left(\frac{3}{8}-3\zeta_2+6\zeta_3\right) -\frac{5}{54} n_f^2 \Bigg\}.   
 \end{align}
The IR singular part of ${\cal S}_f$ ($\langle \mathcal{A}_f| \mathcal{A}_f\rangle$)
computed up to $a_s^5$ agrees perfectly with the predictions based on Catani's proposal. Hence, we present only the finite part of ${\cal S}_f$ for $f=ggg,q\bar{q}g$, namely ${\cal S}_{ggg,fin}$ and
${\cal S}_{q\bar{q}g,fin}$
in the appendix \ref{sec:app}.  

\section{Universality of leading transcendental terms}
\label{sec:LT}
Kotikov and Lipatov ~\cite{Kotikov:2000pm,Kotikov:2002ab,Kotikov:2004er} (see also \cite{Kotikov:2001sc,
Kotikov:2006ts})
conjectured maximum transcendentality principle which relates transcendental terms in
the anomalous dimensions of leading twist two operators in ${\cal N}=4$ SYM with
those in the corresponding QCD results \cite{Moch:2004pa,Vogt:2004mw}.
In the perturbative calculations,
one associates transcendentality weight $n$ to the constants $\zeta(n)$, $\epsilon^{-n}$ and
the functions namely HPLs of weight $n$.
In ${\cal N}=4$ SYM, the FFs of the half-BPS type operators show uniform transcendentality.  On the other hand
QCD results contain terms of all transcendental weights in addition to
rational terms (zero transcendentality).  From various higher order results that are available in QCD and ${\cal N}=4$ SYM theories,
one finds similar connection.  In particular, the leading transcendental terms of anomalous
dimensions of twist two operators, two and three point FFs in QCD when $C_A=C_F=N$ and
$T_f n_f=N/2$ coincide with the corresponding ones of the half-BPS operators in ${\cal N}=4$ SYM.
The leading transcendental terms of
the QCD amplitude for the Higgs boson decaying to three on-shell gluons
~\cite{Gehrmann:2011aa,Koukoutsakis:2003nba} are also related to
the two loop three point MHV FFs of the half-BPS operators~\cite{Brandhuber:2012vm} in ${\cal N}=4$ SYM.
Similar relations are found for two point FFs of quark current, scalar and pseudo-scalar operators constructed
out of gluon field strengths, energy momentum tensor of the QCD up to three loops {see \cite{Ahmed:2015qpa,Ahmed:2016vgl,Ahmed:2016qjf,Gehrmann:2010ue}}.
Recently, in \cite{Banerjee:2016kri}, we studied several three point FFs of the half-BPS as well as the Konishi operators in ${\cal N}=4$ SYM at
two loop level in `t Hooft coupling.  We evaluated them between on-shell final states $g \phi \phi$ and $\phi \lambda \lambda$  
where $\phi,\lambda,g$ are scalar, Majorana and gluon states. Unlike the half-BPS, the Konishi operator is
not protected by supersymmetry and hence we find non-trivial structure at higher orders.  The explicit computation
shows that the FF of the Konishi operator does not show uniform transcendentality~\cite{Ahmed:2016vgl,Banerjee:2016kri,Nandan:2014oga}.  Interestingly, 
the leading transcendental terms of the Konishi for the states 
$g \phi \phi$ agree with those of the half-BPS operator~\cite{Banerjee:2016kri}.  On the other hand, for $\phi \lambda \lambda$ states, we did not
find any such relation to the corresponding ones for the half-BPS.  
In the present context, it is tempting to relate our results for
FFs to the Konishi operator, rather than the half-BPS owing to the fact that both are not protected.  
We find that the amplitude $A \rightarrow g g g$ agrees
with the FF of the Konishi operator between $g \phi \phi $ states at the leading transcendental level while 
$A \rightarrow q \overline q g$ for $O_G$ does not have any relation with that of the Konishi operator computed between $\phi \lambda \lambda$ states.
Surprisingly, we find that the leading transcendental terms of the amplitude $A \rightarrow q \overline q g$ 
multiplied by its born
and normalised by its born square contribution agrees with the half-BPS.  
In summary both the amplitudes namely $A \rightarrow g g g$ and $A \rightarrow q \overline q g$ computed using
$O_G$ operator multiplied by respective 
born amplitudes and normalised by their born squares have leading transcendental terms identical to those of the half-BPS.
Our results with pseudo-scalar operator computed between different on shell states shed more 
light on the universal structure of higher order terms in gauge theories. 
\section{Conclusion}
\label{sec:con}
The motivation to compute two loop virtual corrections to decay of a pseudo-scalar to three gluons or quark, anti-quark, gluon 
is two fold.  Firstly, with larger data set that are available at the LHC, constraining BSM scalar sector 
has become feasible and hence precise theoretical predictions for production of pseudo-scalar and its subsequent decays to
jets need to be achieved.   Secondly, such a computation of NNLO level for the differential rates is technically challenging due to
non-trivial tensorial interaction encountered here.  
We have successfully computed UV renormalised virtual contributions to $A\rightarrow ggg$ and $A\rightarrow q\bar{q}g$ upto two level 
in QCD perturbation series using dimensional regularisation.   The UV renormalisation is performed in ${\overline {MS}}$ scheme. 
We used an effective field theory approach which in our case involves integrating out top quark degrees of freedom.  The resulting effective
operators that contribute to the production of pseudo-scalar contain a Levi-Civita tensor and $\gamma_5$ matrix.   These
objects require special treatment in dimensional regularisation as they can not be defined in arbitrary dimensions other
than four.   Most of prescriptions define them in dimensional regularisation, however all of them 
fail to preserve the symmetries or Ward identities of the theory.   
We have used one of them namely the 't Hooft-Veltman prescription and in order to restore the symmetry we performed a 
non-trivial finite renormalisation 
on top of other standard UV renormalisations.  This way, we could successfully compute UV finite two loop contributions.  
The resulting expressions do contain soft and collinear divergences due to the presence of massless gluons and quarks.
Since these divergences are universal due to factorisation properties of on-shell QCD amplitudes order by order in 
perturbation theory, we confirm the correctness of our results up to finite terms using Catani's
infrared factorisation formula which predict all the poles resulting from soft and collinear configurations up to two loops.   
Expressing our results using Catani's subtraction operators, the remaining infrared finite terms of the amplitudes are 
presented in the appendix \ref{sec:app}.
It is well known in the context of scalar Higgs boson that the leading transcendental terms of finite part of
$H\rightarrow g g g$ amplitude up to two loop level are related to the corresponding ones for three particle amplitude with
an insertion of the half-BPS operator provided the color factors in QCD are adjusted in a specific way.  We find that this is indeed the case
for $A \rightarrow g g g,q \overline q g$ amplitudes presented in this paper.  
In summary, the results for $A \rightarrow ggg$ and $A \rightarrow q \overline q g$  up to two loop level in QCD are presented in this paper.
With appropriate analytical continuation, our results can be readily used for the study of production of a pseudo-scalar in association
with a jet at two loop level at the hadron colliders.

\begin{acknowledgments}
We would like to thank T. Ahmed, G. Das, T. Gehrmann, R. Lee and N. Rana for useful discussions and timely help. We are grateful to Youngjin Kim and Ciaran Williams for their meticulous efforts in comparing our analytical results with theirs~\cite{Kim:2024kaq}.
\end{acknowledgments}

\appendix
\section{Results}
\label{sec:app}
Here, we present finite part of matrix element ${\cal S}_f$ for $f=ggg,q\bar{q}g$, namely ${\cal S}_{ggg,fin}$ and ${\cal S}_{q\bar{q}g,fin}$, by setting $-Q^2=\mu_R^2$ and taking out Born color factor $N(N^2-1)$, $\frac{N^2-1}{2}$ for the final state $g g g$, $q\bar{q}g$ respectively. One loop finite result for $f=ggg$ external state is given by
{\tiny
\begin{align}
S_g^{G,(0)} &= -128\, C_{1, g}\, \left(y^4 + 2 y^3 (-1 + z) + 3 y^2 (-1 + z)^2 + 2 y (-1 + z)^3 +  (1 - z + z^2)^2\right),   \nnn
S_g^{G,(1)} &=
 \bold {N} \Bigg\{ \, 256 \,C_{1, g} \,\left(y^4 + 2 y^3 (-1 + z) + 3 y^2 (-1 + z)^2 + 2 y (-1 + z)^3 +  (1 - z + z^2)^2\right) \Big(\zeta_{2} + H(0; y) (H(0; z) - H(1; z) \nnn &- H(2; y))
+  H(0; z) (H(1; z) - H(2; y)) + 2 H(1; z) H(3; y) + 2 H(1, 0; y) +  2 H(3, 2; y)\Big)
\nnn&
+\frac{1408}{3}C_{1, g} \Big(y^4 + 2 y^3 (-1 + z)  
 + 3 y^2 (-1 + z)^2 + 2 y (-1 + z)^3 +  (1 - z + z^2)^2\Big) \Big(H(0; y) + H(0; z) - H(1; z) - H(2; y)\Big)  
 \nnn&
 -\frac{128}{3}C_{1, g} \Big(24 + 24 y^4  
 + 48 y^3 (-1 + z)  - 47 z + 71 z^2 - 48 z^3 + 24 z^4 +  y (-1 + z)^2 (-47 + 48 z) + y^2 (71 - 143 z + 72 z^2)\Big)\Bigg\} \nnn
&+ \bold {n_f} \Bigg\{\frac{-256 \,C_{1, g}}{3} \Big(y^4 + 2 y^3 (-1 + z) + 3 y^2 (-1 + z)^2 + 2 y (-1 + z)^3 +  (1 - z + z^2)^2\Big) \Big(H(0; y) + H(0; z) - H(1; z) - H(2; y)\Big) \nnn
 & +  \frac{128 \,C_{1, g}}{3}  (-1 + y) (-1 + z) (y + z)
\Bigg\}, \nnn
S_g^{GJ,(1)} & = \bold{n_f} \Bigg\{-64\, C_{1, g}\, \Big(y^4 + 2 y^3 (-1 + z) + 3 y^2 (-1 + z)^2 + 2 y (-1 + z)^3 +  (1 - z + z^2)^2\Big) \Bigg\}.
\end{align}
}
At two loops, for $f=ggg$ external state, we decompose $S_{g}^{G,(2)} $ in terms of the color factors
as follows
\begin{equation}
 S_{g}^{G,(2)} =  S_g^{(2), \, N^2}  + S_g^ {(2), \,n_f/N}+ S_g^{(2), \, n_f  N}+  S_g^{(2), \, n_f^2}    \\
\end{equation} 
where each color term is given below
{\tiny
\begin{align}
S_g^{(2), \, N^2} &=
 \bold{N^2} \, \bigg\{S_g^{G,(0)}\Big( 59 \zeta_{2}^2/10 + 2 \zeta_{3} H(1; z) + 2 \zeta_{3} H(2; y) -  16 \zeta_{2} H(1; z) H(2; y) + 16 H(1; z) H(2; y) H(0, 0; y)\nnn &
-  8 H(1; z) H(3; y) H(0, 0; y) + 8 H(0, 0; y) H(0, 0; z) +  16 \zeta_{2} H(0, 1; y) - 4 H(0, 0; z) H(0, 1; z) + 16 \zeta_{2} H(1, 0; y)\nnn &
+  8 H(1; z) H(3; y) H(1, 0; y) + 8 H(2; y) H(3; y) H(1, 0; y) -  8 H(0, 0; y) H(1, 0; y) - 16 H(2; y) H(3; y) H(1, 0; z)\nnn &
+  16 \zeta_{2} H(1, 1; y) + 8 H(0, 0; y) H(1, 1; z) - 8 H(0, 0; z) H(1, 1; z) -  4 H(0, 0; y) H(2, 0; y) - 16 \zeta_{2} H(2, 2; y)\nnn &
+ 8 H(0, 0; y) H(2, 2; y) +  8 H(0, 0; z) H(2, 2; y) - 16 H(1, 0; z) H(2, 2; y) +  24 H(1; z) H(3; y) H(2, 3; y) - 8 H(1, 0; y) H(2, 3; y)\nnn &
+  24 H(2; y) H(3; y) H(3, 2; y) - 16 H(0, 0; y) H(3, 2; y) -  16 H(1, 0; z) H(3, 3; y) + 32 H(1, 1; z) H(3, 3; y)\nnn &
+  12 H(2; y) H(0, 0, 0; y) + 4 H(1; z) H(0, 0, 1; z) +  16 H(2; y) H(0, 0, 1; z) + 16 H(3; y) H(0, 0, 1; z) -  8 H(1; z) H(0, 0, 2; y)\nnn &
- 24 H(1; y) H(0, 1, 0; y) -  8 H(2; y) H(0, 1, 0; y) + 16 H(2; y) H(0, 1, 1; z) +  16 H(3; y) H(0, 1, 1; z) - 48 H(1; y) H(1, 0, 0; y)\nnn &
-  16 H(2; y) H(1, 0, 0; y) + 16 H(2; y) H(1, 0, 1; z) +  16 H(3; y) H(1, 0, 1; z) + 8 H(1; z) H(2, 0, 0; y) -  8 H(1; z) H(2, 1, 0; y)\nnn &
- 8 H(2; y) H(2, 1, 0; y) +  8 H(1; z) H(2, 2, 3; y) - 24 H(3; y) H(2, 3, 2; y) -  32 H(1; z) H(2, 3, 3; y) + 8 H(1; z) H(3, 0, 0; y)\nnn &
-  2 H(0; z) (\zeta_{3} + 4 H(3; y) H(2, 2; y) +  2 H(2; y) (2 \zeta_{2} - H(0, 0; y) + H(0, 1; z) + H(1, 0; z) + 4 H(1, 1; z) -  H(2, 3; y)\nnn &
+ H(3, 2; y)) - 4 H(1; z) (\zeta_{2} + H(2; y) H(3; y) -  H(0, 0; y) + H(1, 0; y) - H(2, 0; y) + H(2, 2; y) - H(3, 0; y) +  2 H(3, 3; y))\nnn &
- 3 H(0, 0, 1; z) + 2 H(0, 0, 2; y) - 2 H(0, 1, 1; z) -  4 H(0, 2, 2; y) + H(1, 0, 0; z) - 2 H(1, 0, 1; z) - H(1, 1, 1; z) \nnn &+  2 H(2, 0, 0; y)
+ 4 H(2, 1, 0; y) + 4 H(3, 0, 2; y) + 4 H(3, 2, 0; y)) -  48 H(3; y) H(3, 2, 2; y) - 2 H(0; y) (\zeta_{3} + 4 \zeta_{2} H(2; y)
\nnn &+  4 H(2; y) H(0, 0; z)  - 12 H(1; y) H(1, 0; y) + 8 H(3; y) H(1, 1; z) +  H(1; z) (8 H(2; y) H(3; y) - 2 H(0, 0; z) +  4 (\zeta_{2}
+ H(1, 0; y) \nnn &+ H(2, 0; y)  - H(2, 3; y)))  -  2 H(2; y) H(3, 2; y) - 2 H(0; z) (2 \zeta_{2} + 2 H(1; z) H(3; y) +  H(0, 1; z) + 2 H(1, 0; y)
- 2 H(1, 1; z)\nnn & - H(2, 0; y) + 2 H(3, 2; y))  +  2 H(0, 0, 1; z) + H(0, 0, 2; y) - 2 H(1, 0, 0; y) - 2 H(1, 0, 0; z) -  H(2, 0, 0; y)
- H(2, 2, 2; y)\nnn & + 6 H(2, 3, 2; y)  - 4 H(3, 0, 2; y)  -  4 H(3, 2, 0; y) + 12 H(3, 2, 2; y)) + 32 H(1; z) H(3, 3, 2; y)
+  16 H(1; z) H(3, 3, 3; y) \nnn & - 6 H(0, 0, 0, 1; z) - 6 H(0, 0, 0, 2; y)  +  16 H(0, 0, 1, 0; y) + 2 H(0, 0, 1, 0; z) - 6 H(0, 0, 2, 0; y)
+  8 H(0, 0, 3, 2; y) \nnn & + 12 H(0, 1, 0, 0; y)   + 6 H(0, 1, 0, 0; z)  +  4 H(0, 1, 0, 1; z) + 16 H(0, 1, 1, 0; y) + 12 H(0, 1, 1, 0; z)
-  2 H(0, 1, 1, 1; z) \nnn & - 6 H(0, 2, 0, 0; y)  - 2 H(0, 2, 2, 2; y)   +  12 H(1, 0, 0, 0; y)  + 6 H(1, 0, 0, 0; z) + 4 H(1, 0, 0, 1; z)
+  24 H(1, 0, 1, 0; y) \nnn & + 12 H(1, 0, 1, 0; z)  - 2 H(1, 0, 1, 1; z)  +  32 H(1, 1, 0, 0; y) + 16 H(1, 1, 0, 0; z) - 2 H(1, 1, 0, 1; z)
+  16 H(1, 1, 1, 0; y) \nnn &- 2 H(1, 1, 1, 0; z)  - 6 H(2, 0, 0, 0; y) -  2 H(2, 0, 2, 2; y) - 2 H(2, 2, 0, 2; y)  + 8 H(2, 2, 1, 0; y)
 -  2 H(2, 2, 2, 0; y)\nnn &  + 8 H(2, 2, 3, 2; y)  + 16 H(2, 3, 3, 2; y) +  24 H(3, 2, 3, 2; y) + 32 H(3, 3, 2, 2; y) + 16 H(3, 3, 3, 2; y)\Big) \nnn
 &+ C_{1, g}\Big( (64 (33 + 121 y^4 - 114 z + 231 z^2 - 194 z^3 + 121 z^4 +  2 y^3 (-97 + 121 z) + 3 y^2 (77 - 202 z + 121 z^2) +  2 y (-57 + 243 z\nnn &
- 303 z^2 + 121 z^3)) \zeta_{3})/3 -  (64 (121 y^4 + 218 y^3 (-1 + z) + 297 y^2 (-1 + z)^2 + 178 y (-1 + z)^3 +  77 (1 - z
+ z^2)^2) \zeta_{2} H(0; y))/3 
\nnn &
 -  (64 (77 + 77 y^4 - 178 z + 297 z^2 - 218 z^3 + 121 z^4 +  2 y^3 (-77 + 89 z) + 3 y^2 (77 - 178 z + 99 z^2)\nnn &
+  2 y (-77 + 267 z - 297 z^2 + 109 z^3)) \zeta_{2} H(0; z))/3 -  (64 (11 + 99 y^4 + 74 z - 111 z^2 + 114 z^3 - 33 z^4 + y^3 (-62 + 54 z)\nnn &
+  3 y^2 (11 - 14 z + 7 z^2) + y (18 - 78 z + 78 z^2 - 66 z^3)) \zeta_{2}  H(1; z))/3 - (64 (55 + 11 y^4 + 74 z - 111 z^2 + 114 z^3 + 11 z^4\nnn &
-  2 y^3 (-57 + 61 z) - 3 y^2 (37 - 82 z + 41 z^2) +  y (74 - 246 z + 246 z^2 - 122 z^3)) \zeta_{2} H(2; y))/3 -  (128 (22 y^4 + 32 y^3 (-1\nnn &
+ z) + 63 y^2 (-1 + z)^2 + 42 y (-1 + z)^3 +  11 (3 - 4 z + 6 z^2 - 4 z^3 + 3 z^4)) H(0; z) H(1; z) H(2; y))/3 +  (128 (55 y^4 + y^3 (-88\nnn &
+ 74 z) + 6 y^2 (22 - 44 z + 21 z^2) +  y (-88 + 264 z - 264 z^2 + 74 z^3) +  11 (4 - 8 z + 12 z^2 - 8 z^3 + 5 z^4)) H(1; z) H(3; y)^2)/3\nnn &
+  (128 (77 + 55 y^4 - 86 z + 129 z^2 - 76 z^3 + 66 z^4 + y^3 (-64 + 62 z) +  3 y^2 (42 - 82 z + 41 z^2) + 6 y (-14 + 41 z - 41 z^2\nnn &
+ 12 z^3)) H(2; y)  H(0, 1; z))/3 - (128 y (22 y^3 + 32 y^2 (-1 + z) + 33 y (-1 + z)^2 +  12 (-1 + z)^3) (H(0; y) H(0; z) H(1; z)\nnn &
- H(0; y) H(0, 1; z)))/3 -  (128 z (-12 + 12 y^3 + 33 z - 32 z^2 + 22 z^3 + y^2 (-36 + 33 z) +  y (36 - 66 z \nnn & + 32 z^2)) H(0; z) H(1, 0; y))/3
+  (128 (33 + 33 y^4 - 90 z + 135 z^2 - 100 z^3 + 44 z^4 + y^3 (-88 + 90 z) +  3 y^2 (44 - 90 z + 45 z^2) \nnn & + 2 y (-44 + 135 z - 135 z^2
+ 50 z^3))  (H(1; z) H(1, 0; y) + H(2; y) H(1, 0; y)))/3 +  (128 (22 + 33 y^4 - 90 z + 135 z^2 - 100 z^3  \nnn &+ 33 z^4 +  2 y^3 (-50 + 51 z)
+ 3 y^2 (45 - 92 z + 46 z^2) +  6 y (-15 + 46 z - 46 z^2 + 17 z^3)) H(1; z) H(3, 2; y))/3 +  (128 z (12 y^3  \nnn & + 3 y^2 z + 2 y z^2
- 11 z^3) (H(0; z) H(1; z) H(3; y) +  H(0; z) H(3, 2; y)))/3 - (128 y (11 y^3 - 2 y^2 z - 3 y z^2  \nnn & - 12 z^3)  (H(0; y) H(1; z) H(3; y)
+ H(0; y) H(0, 1; z) + H(0; y) H(3, 2; y) -  2 H(0, 0, 1; z)))/3 -  (128 (99 y^4 + y^3 (-232 + 234 z) \nnn & + 132 (1 - z + z^2)^2 +  y^2 (363 
- 726 z + 366 z^2) + 12 y (-21 + 63 z - 63 z^2 + 22 z^3))  H(0, 1, 0; y))/3 - (128 (88 + 88 y^4 - 164 z  \nnn &+ 231 z^2 - 144 z^3 +  66 z^4
+ 4 y^3 (-44 + 41 z) + 3 y^2 (88 - 164 z + 77 z^2) +  2 y (-88 + 246 z - 231 z^2 + 72 z^3)) H(0, 1, 0; z))/3  \nnn &+  (128 (77 + 77 y^4 - 130 z
+ 195 z^2 - 120 z^3 + 66 z^4 +  2 y^3 (-66 + 65 z) + 3 y^2 (66 - 130 z + 65 z^2) +  6 y (-22 + 65 z - 65 z^2   \nnn & + 20 z^3)) H(1, 0, 1; z))/3
+  (256 (11 + 11 y^4 + 14 z - 14 y^3 z - 36 z^2 + 44 z^3 - 22 z^4 -  6 y^2 z (-7 + 6 z) - 2 y z (21 - 36 z + 22 z^2)) \nnn &  (\zeta_{2} H(1; y)
+ H(1, 1, 0; y)))/3  -  (256 (11 y^4 + 10 y^3 (-1 + z) - 10 y (-1 + z)^3 - 11 (1 - z + z^2)^2)  H(1, 1, 0; z))/3 + (128 (110  \nnn & + 99 y^4 - 174 z
+ 261 z^2 - 164 z^3 +  99 z^4 + 2 y^3 (-82 + 81 z) + 3 y^2 (87 - 172 z + 86 z^2) +  6 y (-29 + 86 z - 86 z^2
 \nnn &+ 27 z^3)) (H(1; z) H(2; y) H(3; y)  +  H(2, 3, 2; y)))/3 - 2816 (y^4 + 2 y^3 (-1 + z) + 3 y^2 (-1 + z)^2 +  2 y (-1 + z)^3 \nnn & + (1 - z
+ z^2)^2) (-(H(0; y) H(0; z) H(2; y))/3  +  H(0; y) H(1; z) H(2; y) + (H(0; y)^2 H(0; z) + H(0; y) H(0; z)^2 \nnn & -  H(0; y)^2 H(1; z)
+ H(0; y) H(1; z)^2 - H(0; z)^2 H(2; y)  +  H(0; z) H(2; y)^2)/2 + (H(0; y) H(1; z) H(3; y))/3 \nnn & +  (H(0; z) H(1; z) H(3; y))/3 
- H(1; z)^2 H(3; y) + H(1; z) H(3; y)^2  -  H(0; y) H(0, 2; y) + (H(0; z) H(1, 0; y))/3  \nnn &+ H(0; y) H(2, 2; y) +  (H(0; y) H(3, 2; y))/3
+ (H(0; z) H(3, 2; y))/3 + H(0, 0, 1; z)/3 +  H(0, 0, 2; y) - H(0, 1, 1; z) \nnn & + 2 H(1, 0, 0; y) + H(1, 0, 0; z) -  H(2, 0, 0; y)
- 2 H(3, 2, 2; y)) -  (256 (11 y^4 + y^3 (-44 + 58 z) + 6 y^2 (11 - 22 z + 12 z^2)  \nnn &+  2 y (-22 + 66 z - 66 z^2 + 29 z^3) + 11 (2 - 4 z
+ 6 z^2 - 4 z^3 + z^4))  (H(3; y) H(0, 1; z) + H(3, 3, 2; y)))/3 \Big) \nnn
  &+ C_{3, g} \Big( ((-128 y^2 z^2 (64 y^6 + 2 y^5 (-125 + 62 z) +  y^4 (506 - 691 z + 149 z^2) + 2 y^3 (-329 + 771 z - 498 z^2 + 59 z^3) +  2 (-1\nnn &
+ z)^2 (32 - 70 z + 99 z^2 - 61 z^3 + 32 z^4) +  y^2 (542 - 1797 z + 2076 z^2 - 996 z^3 + 149 z^4) +  y (-268 + 1090 z - 1797 z^2 \nnn &+ 1542 z^3
- 691 z^4 + 124 z^5)) \zeta_{2})/9 +  (128 y^2 z^2 (157 y^6 + y^5 (-628 + 760 z) +  y^4 (1250 - 2851 z + 1619 z^2) + 2 y^3 (-776 \nnn &+ 2574 z 
- 2841 z^2 + 1046 z^3)  + (-1 + z)^2 (157 - 308 z + 465 z^2 - 314 z^3 + 157 z^4) +  y^2 (1238 - 4941 z + 8013 z^2 - 5682 z^3\nnn & + 1619 z^4)
+  y (-622 + 2506 z - 4941 z^2 + 5148 z^3 - 2851 z^4 + 760 z^5)) H(0; y)  H(0; z))/9 + (128 y^2 z^2 (133 y^6 + y^5 (-532 \nnn &+ 576 z)
+  y^4 (1060 - 2243 z + 1185 z^2) + 2 y^3 (-659 + 2068 z - 2160 z^2 +  752 z^3) + (-1 + z)^2 (133 - 266 z + 401 z^2 - 268 z^3\nnn & + 133 z^4)
+  y^2 (1052 - 4063 z + 6311 z^2 - 4322 z^3 + 1185 z^4) +  y (-528 + 2126 z - 4079 z^2 + 4152 z^3 - 2247 z^4 \nnn & + 576 z^5)) H(0; z)  H(1; z))/3 
+ (128 z^2 (-1 + y + z)^2 (-157 y^6 + y^5 (314 - 182 z) +  60 (-1 + z)^2 z^2 (1 - z + z^2) - 3 y^4 (155 \nnn & - 207 z + 58 z^2) +  2 y^3 (154
- 258 z + 69 z^2 + 38 z^3) +  15 y z (10 - 37 z + 58 z^2 - 43 z^3 + 12 z^4) +  y^2 (-157 - 73 z + 561 z^2  \nnn &- 766 z^3
+ 188 z^4)) H(0; z) H(2; y))/9  +  (128 y^2 (-1 + y + z)^2 (60 y^6 + 180 y^5 (-1 + z) +  y^4 (240 - 645 z + 188 z^2) \nnn &+ 2 y^3 (-90 + 435 z
- 383 z^2 + 38 z^3)  +  y z (150 - 73 z - 516 z^2 + 621 z^3 - 182 z^4) +  y^2 (60 - 555 z + 561 z^2  + 138 z^3 - 174 z^4)\nnn & +  z^2 (-157 + 308 z 
- 465 z^2 + 314 z^3 - 157 z^4))  (H(0; y) H(1; z) + H(0; y) H(2; y)))/9 \nnn & -  (15488 y^2 z^2 (-1 + y + z)^2 (y^4 + 2 y^3 (-1 + z) + 3 y^2 (-1
+ z)^2 +  2 y (-1 + z)^3 + (1 - z + z^2)^2) (H(0; y)^2 + H(0; z)^2\nnn & + H(1; z)^2+  2 H(1; z) H(2; y) + H(2; y)^2))/9 -  (128 y^2 (20 y^8
+ 100 y^7 (-1 + z) - 2 (-1 + z)^3 z^3 (-1 + 2 z) \nnn &+  5 y^6 (44 - 99 z + 51 z^2) + 2 y^5 (-140 + 530 z - 601 z^2 + 232 z^3) +  y^4 (220
- 1260 z + 2271 z^2 - 1816 z^3 + 595 z^4)\nnn & +  y^3 (-100 + 870 z - 2181 z^2 + 2668 z^3 - 1805 z^4 + 546 z^5) +  y z (50 - 225 z + 452 z^2
- 625 z^3 + 582 z^4 - 322 z^5 + 88 z^6) \nnn &+  y^2 (20 - 325 z + 1082 z^2 - 1766 z^3 + 1825 z^4 - 1114 z^5 + 308 z^6))  H(0, 1; z))/3 
- (128 y^2 (60 y^8 + 300 y^7 (-1 + z) +  11 y^6 (60\nnn & - 135 z + 19 z^2) - 2 y^5 (420 - 1590 z + 685 z^2 +  548 z^3) + y^4 (660 - 3780 z
+ 2341 z^2 + 4174 z^3 - 3389 z^4) -  2 (-1 + z)^2 z^2 (278\nnn & - 547 z + 825 z^2 - 556 z^3 + 278 z^4) -  y^3 (300 - 2610 z + 983 z^2 + 9648 z^3
- 13293 z^4 + 4966 z^5) +  y^2 (60 - 975 z - 1178 z^2 \nnn & + 11892 z^3 - 21561 z^4 + 15366 z^5 -  4250 z^6) + y z (150 + 1537 z - 7528 z^2\nnn &
+ 15315 z^3 - 15906 z^4 +  8656 z^5 - 2224 z^6)) H(1, 0; y))/9 -  (128 (-1 + y + z)^2 (60 y^8 + 180 y^7 (-1 + z) +  y^6 (240 - 645 z\nnn &
- 211 z^2) + 60 (-1 + z)^2 z^4 (1 - z + z^2) +  y^5 (-180 + 870 z + 26 z^2 - 590 z^3) +  y^3 z (150 + 719 z - 2484 z^2 + 2199 z^3 - 590 z^4)\nnn &
+  y^2 z^2 (-556 + 719 z - 624 z^2 + 26 z^3 - 211 z^4) +  15 y z^3 (10 - 37 z + 58 z^2 - 43 z^3 + 12 z^4) -  3 y^4 (-20 + 185 z + 208 z^2\nnn &
- 733 z^3 + 358 z^4))  (H(1; z) H(3; y) + H(3, 2; y)))/9)\Big) \nnn
&+ C_{2, g} \Big(\, (128 (-1 + y) y (-1 + z)^2 (y + z)^2 (60 y^7 + 15 y^6 (-16 + 39 z) +  2 y^5 (210 - 1209 z + 841 z^2) + y^4 (-420 + 4639 z \nnn & - 7039 z^2
+  2851 z^3) + y^3 (240 - 5265 z + 12582 z^2 - 10570 z^3 + 3068 z^4) +  z (381 - 1121 z + 1861 z^2 - 1883 z^3 + 1143 z^4 \nnn & - 381 z^5) 
+  y z (-1663 + 5417 z - 8590 z^2 + 6912 z^3 - 3114 z^4 + 657 z^5) +  y^2 (-60 + 3741 z - 11521 z^2 + 14448 z^3 - 8097 z^4\nnn &
+ 1971 z^5))  H(0; y))/9 + (128 (-1 + y)^2 (-1 + z) z (y + z)^2  (y^6 (-381 + 657 z) + 60 (-1 + z)^3 z^2 (1 - z + z^2) +  9 y^5 (127 \nnn & - 346 z
+ 219 z^2) + y^3 (-1 + z)^2  (1861 - 4868 z + 2851 z^2) + y^4 (-1883 + 6912 z - 8097 z^2 +  3068 z^3) + y (-1 + z)^2 (381\nnn & - 901 z  + 1558 z^2
- 1248 z^3 +  585 z^4) + y^2 (-1121 + 5417 z - 11521 z^2 + 12582 z^3 - 7039 z^4 +  1682 z^5)) H(0; z))/9 \nnn & + (128 (-1 + y)^2 (-1
+ z)^2  (60 y^9 - 15 y^8 (16 + 3 z) + y^7 (420 - 243 z - 838 z^2) +  60 (-1 + z)^3 z^4 (1 - z + z^2) + y^6 (-420\nnn & + 524 z + 1966 z^2
-  2417 z^3) + y^5 (240 - 385 z - 3059 z^2 + 7144 z^3 - 3916 z^4) +  y^4 (-60 + 14 z + 2318 z^2 - 9133 z^3 \nnn & + 10350 z^4 - 3916 z^5)
+  y^3 z (135 - 1158 z + 4566 z^2 - 9133 z^3 + 7144 z^4 - 2417 z^5) +  y^2 z^2 (390 - 1158 z + 2318 z^2 - 3059 z^3\nnn & + 1966 z^4 - 838 z^5)
+  y z^3 (135 + 14 z - 385 z^2 + 524 z^3 - 243 z^4 - 45 z^5))  (H(1; z) + H(2; y)))/9 
\Big) \nnn
&- \frac{128 \, C_{1, g}}{27} \Big(  3256 + 3169 y^4 + 6338 y^3 (-1 + z) - 6148 z + 9317 z^2 -  6338 z^3 + 3169 z^4 + y^2 (9317 - 18731 z + 9507 z^2)\nnn & +  y (-6148 + 18541 z - 18731 z^2 + 6338 z^3)
\Big)\bigg\}.\\
S_g^ {(2),\,n_f/N}&=
  \bold{\frac{n_f}{N}} \Bigg\{  1024 \,C_{1, g} \Big(y^4 + 2 y^3 (-1 + z) + 3 y^2 (-1 + z)^2 + 2 y (-1 + z)^3 +  (1 - z + z^2)^2\Big) \zeta_{3}
 \nnn
& + \frac{128 \,C_{3, g}}{3}\Big( y^3 z^3 (6 + 7 y^4 - 25 z + 35 z^2 - 23 z^3 + 7 z^4 +  y^3 (-23 + 20 z) + y^2 (35 - 57 z + 24 z^2) +  y (-25 + 68 z - 57 z^2\nnn &
+ 20 z^3)) (\zeta_{2} + H(0; y) H(0; z) +  H(0; z) H(1; z)) +  z^3 (-1 + y + z)^3  (7 y^4 + y^3 (-5 + 8 z) + 2 y^2 (4 - 6 z + 3 z^2)\nnn &
+  y (-4 + 2 z - 4 z^3) - 2 z (-1 + 2 z - 2 z^2 + z^3)) H(0; z) H(2; y) -  y^3 (-1 + y + z)^3 (2 y^4 + 4 y^3 (-1 + z) + y^2 (4
- 6 z^2) \nnn & +  z (4 - 8 z + 5 z^2 - 7 z^3) - 2 y (1 + z - 6 z^2 + 4 z^3))  (H(0; y) H(1; z) + H(0; y) H(2; y))+   y^3 (2 y^7
+ 10 y^6 (-1 + z) \nnn & + 2 y^5 (11 - 18 z + 6 z^2) -  y^4 (28 - 52 z + 12 z^2 + 19 z^3) + y^3 (22 - 36 z - 32 z^2 + 108 z^3 -  65 z^4) + y^2 (-10+ 6 z + 72 z^2 \nnn & - 190 z^3 + 195 z^4 - 75 z^5) +  y (2 + 8 z - 60 z^2 + 148 z^3 - 205 z^4 + 150 z^5 - 49 z^6) +  z (-4 + 20 z - 47 z^2 + 75 z^3
- 79 z^4 \nnn &+ 49 z^5  - 14 z^6)) H(0, 1; z) +  y^3 (2 y^7 + 10 y^6 (-1 + z) + 2 y^5 (11 - 18 z + 6 z^2) -  y^4 (28 - 52 z + 12 z^2
+ 5 z^3) + y^3 (22 - 36 z \nnn & - 32 z^2  + 62 z^3  -  25 z^4) + y^2 (-10 + 6 z + 72 z^2 - 120 z^3 + 81 z^4 - 27 z^5) +  z (-4 + 20 z - 35 z^2
+ 25 z^3 - 9 z^4 + 3 z^5)  +  y (2 \nnn & + 8 z   - 60 z^2 + 98 z^3  - 69 z^4  + 36 z^5 - 9 z^6)) H(1, 0; y) +   (-1 + y + z)^3 (2 y^7 + 4 y^6 (-1
+ z) + y^5 (4 - 6 z^2)  -  2 y^2 z^3 (4 - 6 z \nnn &+ 3 z^2)+ y^3 z (4 - 8 z  + 10 z^2 - 15 z^3) +  2 z^4 (-1 + 2 z - 2 z^2 + z^3) + 2 y z^3 (2 - z
+ 2 z^3) -  y^4 (2 + 2 z - 12 z^2 \nnn &+ 15 z^3)) (H(1; z) H(3; y)
+ H(3, 2; y))
\Big) \nnn
& +  C_{4, g} \Big((-128 y^2 (-1 + z) (2 y^7 + y^6 (-8 + 9 z) + y^5 (14 - 25 z - 2 z^2) +  y^4 (-14 + 28 z + 37 z^2 - 59 z^3) +  y^3 (8 - 15 z - 78 z^2\nnn &
+ 204 z^3 - 120 z^4) +  y^2 (-2 + z + 66 z^2 - 242 z^3 + 290 z^4 - 124 z^5) +  y z (2 - 27 z + 117 z^2 - 220 z^3 + 194 z^4 - 72 z^5)\nnn &
+  2 z^2 (2 - 10 z + 25 z^2 - 35 z^3 + 27 z^4 - 9 z^5)) H(0; y))/3 +  (128 (-1 + y) z^2 (18 y^7 + 18 y^6 (-3 + 4 z) +  10 y^4 (-1 + z)^2 (-5\nnn &
+ 12 z) - 2 (-1 + z)^3 z^2 (1 - z + z^2) +  2 y^5 (35 - 97 z + 62 z^2) - y (-1 + z)^2 z (2 + 5 z - 7 z^2 + 9 z^3) +  y^3 (20 - 117 z\nnn &
+ 242 z^2 - 204 z^3 + 59 z^4) +  y^2 (-4 + 27 z - 66 z^2 + 78 z^3 - 37 z^4 + 2 z^5)) H(0; z))/3 -  (128 (-1 + y) (-1 + z) (-1 + y\nnn &
+ z)^2 (2 y^6 + y^5 (-4 + 5 z) -  2 y^4 (-2 + z + 7 z^2) + y^2 z (3 - 8 z + 14 z^2 - 14 z^3) +  2 z^3 (-1 + 2 z - 2 z^2 + z^3)\nnn & + y z^2 (3
- 2 z^2 + 5 z^3) -  2 y^3 (1 - 7 z^2 + 8 z^3)) (H(1; z) + H(2; y)))/3
\Big) \nnn
& - \frac{64 \, C_{1, g}}{3} \Big((142 + 144 y^4 + 288 y^3 (-1 + z) - 289 z + 433 z^2 - 288 z^3 +  144 z^4 + y^2 (433 - 867 z + 432 z^2) +  y (-289 + 868 z  \nnn & - 867 z^2 + 288 z^3))
\Big) \Bigg\}.\\
S_g^{(2), \, n_f  N} &=
  \bold{n_f N} \Bigg\{ C_{1, g}\Big(  (128 (21 + 13 y^4 - 36 z + 51 z^2 - 32 z^3 + 13 z^4 + y^3 (-32 + 26 z) +  y^2 (51 - 84 z + 39 z^2) + y (-36 + 90 z - 84 z^2\nnn &
+ 26 z^3)) \zeta_{3})/3 +  (128 (11 y^4 + 19 y^3 (-1 + z) + 27 y^2 (-1 + z)^2 + 17 y (-1 + z)^3 +  7 (1 - z + z^2)^2) \zeta_{2} H(0; y))/3\nnn &
+  (128 (7 + 7 y^4 - 17 z + 27 z^2 - 19 z^3 + 11 z^4 + y^3 (-14 + 17 z) +  3 y^2 (7 - 17 z + 9 z^2) + y (-14 + 51 z - 54 z^2\nnn &
+ 19 z^3)) \zeta_{2}  H(0; z))/3 + (128 (1 - 4 y^3 + 9 y^4 + 10 z - 15 z^2 + 12 z^3 - 3 z^4 +  y^2 (3 + 6 z - 3 z^2) - 6 y z (2 - 2 z\nnn &
+ z^2)) \zeta_{2} H(1; z))/3 +  (128 (5 + y^4 + 10 z - 15 z^2 + 12 z^3 + z^4 - 4 y^3 (-3 + 4 z) -  3 y^2 (5 - 14 z + 7 z^2) - 2 y (-5 + 21 z\nnn &
- 21 z^2 + 8 z^3)) \zeta_{2}  H(2; y))/3 + (128 (6 + 4 y^4 + 5 y^3 (-1 + z) + 9 y^2 (-1 + z)^2 +  6 y (-1 + z)^3 - 8 z + 12 z^2 - 8 z^3\nnn &
+ 6 z^4) H(0; z) H(1; z) H(2; y))/ 3 + (128 (2 y^4 + y^3 (-8 + 13 z) + 6 y^2 (2 - 4 z + 3 z^2) +  y (-8 + 24 z - 24 z^2 + 13 z^3) + 2 (2\nnn &
- 4 z + 6 z^2 - 4 z^3 + z^4))  H(1; z) H(3; y)^2)/3 + (128 y (2 y^3 - 2 y^2 z - 3 y z^2 - 3 z^3) H(0; y)  H(0, 1; z))/3 - (128 (14 + 10 y^4\nnn &
- 14 z + 21 z^2 - 13 z^3 + 12 z^4 +  2 y^3 (-5 + 4 z) + 3 y^2 (6 - 10 z + 5 z^2) +  3 y (-4 + 10 z - 10 z^2 + 3 z^3)) H(2; y) H(0, 1; z))/3\nnn &
+  (128 y (4 y^3 + 5 y^2 (-1 + z) + 6 y (-1 + z)^2 + 3 (-1 + z)^3)  (H(0; y) H(0; z) H(1; z) - H(0; y) H(0, 1; z)))/3 +  (128 (4 + 4 y^4\nnn &
- 11 z + 18 z^2 - 13 z^3 + 8 z^4 + y^3 (-8 + 11 z) +  3 y^2 (4 - 11 z + 6 z^2) + y (-8 + 33 z - 36 z^2 + 13 z^3)) H(0; z)  H(1, 0; y))/3\nnn &
- (128 (6 + 6 y^4 - 18 z + 27 z^2 - 19 z^3 + 8 z^4 +  2 y^3 (-8 + 9 z) + 3 y^2 (8 - 18 z + 9 z^2) +  y (-16 + 54 z - 54 z^2\nnn &
+ 19 z^3)) (H(1; z) H(1, 0; y) +  H(2; y) H(1, 0; y)))/3 -  (128 (4 + 6 y^4 - 18 z + 27 z^2 - 19 z^3 + 6 z^4 + y^3 (-19 + 21 z) +  3 y^2 (9\nnn &
- 20 z + 10 z^2) + 3 y (-6 + 20 z - 20 z^2 + 7 z^3)) H(1; z)  H(3, 2; y))/3 + (128 (6 y^4 + y^3 (-8 + 6 z) + 4 (1 - z + z^2)^2 +  3 y^2 (4\nnn &
- 8 z + 3 z^2) + y (-8 + 24 z - 24 z^2 + 5 z^3))  (H(0; y) H(1; z) H(3; y) + H(0; y) H(3, 2; y)))/3 +  (128 (4 + 4 y^4 - 8 z + 12 z^2 \nnn & - 8 z^3
+ 6 z^4 + y^3 (-8 + 5 z) +  3 y^2 (4 - 8 z + 3 z^2) + y (-8 + 24 z - 24 z^2 + 6 z^3))  (H(0; z) H(1; z) H(3; y)\nnn & + H(0; z) H(3, 2; y)))/3
+  (256 (y^3 (-4 + 6 z) + 2 (1 - z + z^2)^2 + 3 y^2 (2 - 4 z + 3 z^2) +  y (-4 + 12 z\nnn & - 12 z^2 + 7 z^3)) H(0, 0, 1; z))/3 +  (128 (18 y^4
+ y^3 (-43 + 45 z) + 24 (1 - z + z^2)^2 \nnn & +  3 y^2 (22 - 44 z + 23 z^2) + 3 y (-15 + 45 z - 45 z^2 + 16 z^3))  H(0, 1, 0; y))/3 + (128 (16
+ 16 y^4 - 29 z \nnn &+ 42 z^2 - 27 z^3 + 12 z^4 +  y^3 (-32 + 29 z) + y^2 (48 - 87 z + 42 z^2) +  y (-32 + 87 z - 84 z^2\nnn &
+ 27 z^3)) H(0, 1, 0; z))/3 -  (128 (14 + 14 y^4 - 22 z + 33 z^2 - 21 z^3 + 12 z^4 + y^3 (-24 + 22 z) +  y^2 (36 - 66 z + 33 z^2) + 3 y (-8\nnn &
+ 22 z - 22 z^2 + 7 z^3))  H(1, 0, 1; z))/3 - (256 (2 + 2 y^4 + 5 z - 5 y^3 z + 3 y^2 (5 - 3 z) z -  9 z^2 + 8 z^3 - 4 z^4 + y z (-15 + 18 z\nnn &
- 8 z^2))  (\zeta_{2} H(1; y) + H(1, 1, 0; y)))/3 +  (256 (2 y^4 + y^3 (-1 + z) - y (-1 + z)^3 - 2 (1 - z + z^2)^2)  H(1, 1, 0; z))/3\nnn &
- (128 (20 + 18 y^4 - 30 z + 45 z^2 - 29 z^3 + 18 z^4 +  y^3 (-29 + 27 z) + y^2 (45 - 84 z + 42 z^2) +  3 y (-10 + 28 z - 28 z^2\nnn &
+ 9 z^3)) (H(1; z) H(2; y) H(3; y) +  H(2, 3, 2; y)))/3 + 512 (y^4 + 2 y^3 (-1 + z) + 3 y^2 (-1 + z)^2 +  2 y (-1 + z)^3 + (1 - z\nnn &
+ z^2)^2) ((H(0; y)^2 H(0; z))/2 +  (H(0; y) H(0; z)^2)/2 - (H(0; y)^2 H(1; z))/2 + (H(0; y) H(1; z)^2)/2\nnn &  -  (H(0; y) H(0; z) H(2; y))/3
- (H(0; z)^2 H(2; y))/2 +  H(0; y) H(1; z) H(2; y) + (H(0; z) H(2; y)^2)/2 - H(1; z)^2 H(3; y) \nnn &-  H(0; y) H(0, 2; y)  + H(0; y) H(2, 2; y)
+ H(0, 0, 2; y) - H(0, 1, 1; z) +  2 H(1, 0, 0; y) + H(1, 0, 0; z) - H(2, 0, 0; y)\nnn & - 2 H(3, 2, 2; y)) +  (256 (2 y^4  + y^3 (-8 + 13 z)
+ 6 y^2 (2 - 4 z + 3 z^2) +  y (-8 + 24 z - 24 z^2 + 13 z^3)\nnn & + 2 (2 - 4 z + 6 z^2 - 4 z^3 + z^4))  (H(3; y) H(0, 1; z) + H(3, 3, 2; y)))/3
\Big) \nnn
& + C_{3, g} \Big( (64 y^2 z^2 (101 y^6 + y^5 (-392 + 290 z) +  2 y^4 (398 - 679 z + 236 z^2) + 2 y^3 (-523 + 1404 z - 1086 z^2 +  217 z^3) + (-1\nnn &
+ z)^2 (101 - 226 z + 315 z^2 - 190 z^3 + 101 z^4) +  y^2 (868 - 3096 z + 3993 z^2 - 2172 z^3 + 472 z^4) +  2 y (-214 + 892 z - 1548 z^2\nnn &
+ 1404 z^3 - 679 z^4 + 145 z^5)) \zeta_{2})/9 +  (128 y^2 z^2 (8 y^6 + y^5 (-32 + 13 z) + y^4 (66 - 83 z + 8 z^2) -  y^3 (86 - 183 z + 81 z^2\nnn &
+ 16 z^3) + 2 (-1 + z)^2  (4 - 9 z + 13 z^2 - 8 z^3 + 4 z^4) +  y^2 (70 - 215 z + 220 z^2 - 81 z^3 + 8 z^4) +  y (-34 + 136 z - 215 z^2\nnn &
+ 183 z^3 - 83 z^4 + 13 z^5)) H(0; y) H(0; z))/ 3 - (128 y^2 z^2 (64 y^6 + y^5 (-256 + 313 z) +  5 y^4 (100 - 233 z + 136 z^2) + y^3 (-604\nnn &
+ 2091 z - 2415 z^2 +  928 z^3) + 2 (-1 + z)^2 (32 - 64 z + 99 z^2 - 67 z^3 + 32 z^4) +  y^2 (476 - 1983 z + 3330 z^2 - 2421 z^3 + 680 z^4)  
\nnn &
+  y (-244 + 1000 z - 2031 z^2 + 2139 z^3 - 1177 z^4 + 313 z^5)) H(0; z)  H(1; z))/9 - (128 z^2 (-1 + y + z)^2 (8 y^6 + y^5 (-16 + 35 z)\nnn &
+  22 (-1 + z)^2 z^2 (1 - z + z^2) + y^4 (26 - 98 z + 63 z^2) +  3 y^3 (-6 + 41 z - 61 z^2 + 26 z^3) +  2 y z (14 - 68 z + 119 z^2 - 98 z^3\nnn &
+ 33 z^4) +  y^2 (8 - 88 z + 238 z^2 - 250 z^3 + 94 z^4)) H(0; z) H(2; y))/3 -  (128 y^2 (-1 + y + z)^2 (22 y^6 + 66 y^5 (-1 + z)\nnn &
+  2 y^4 (44 - 98 z + 47 z^2) + y^3 (-66 + 238 z - 250 z^2 + 78 z^3) +  2 z^2 (4 - 9 z + 13 z^2 - 8 z^3 + 4 z^4) +  y z (28 - 88 z + 123 z^2\nnn &
- 98 z^3 + 35 z^4) +  y^2 (22 - 136 z + 238 z^2 - 183 z^3 + 63 z^4))  (H(0; y) H(1; z) + H(0; y) H(2; y)))/3 +  (5632 y^2 z^2 (-1 + y\nnn &
+ z)^2 (y^4 + 2 y^3 (-1 + z) + 3 y^2 (-1 + z)^2 +  2 y (-1 + z)^3 + (1 - z + z^2)^2) (H(0; y)^2 + H(0; z)^2 + H(1; z)^2\nnn & +  2 H(1; z) H(2; y)
+ H(2; y)^2))/9 +  (128 y^2 (22 y^8 + 110 y^7 (-1 + z) - 2 (-1 + z)^3 z^3 (-1 + 2 z) +  y^6 (242 - 504 z + 240 z^2) \nnn &+ y^5 (-308 + 1004 z
- 998 z^2 + 319 z^3) +  y^4 (242 - 1116 z + 1726 z^2 - 1150 z^3 + 305 z^4) \nnn &+  y^3 (-110 + 714 z - 1542 z^2 + 1614 z^3 - 935 z^4 + 255 z^5)
+  y z (28 - 124 z + 243 z^2 - 325 z^3 \nnn &+ 287 z^4 - 147 z^5 + 38 z^6) +  y^2 (22 - 236 z + 698 z^2 - 1024 z^3 + 945 z^4 - 532 z^5
+ 133 z^6))  H(0, 1; z))/3 \nnn &+ (128 y^2 (66 y^8 + 330 y^7 (-1 + z) +  2 y^6 (363 - 756 z + 340 z^2) + y^5 (-924 + 3012 z - 2822 z^2 +  683 z^3)
\nnn &
- 2 (-1 + z)^2 z^2 (20 - 31 z + 51 z^2 - 40 z^3 + 20 z^4) +  y^4 (726 - 3348 z + 4834 z^2 - 2492 z^3 + 259 z^4) -  y^3 (330 - 2142 z\nnn &
+ 4226 z^2 - 3204 z^3 + 567 z^4 + 211 z^5) +  y^2 (66 - 708 z + 1798 z^2 - 1674 z^3 + 75 z^4 + 642 z^5 - 257 z^6) +  y z (84 - 224 z\nnn &
+ 137 z^2 + 423 z^3 - 777 z^4 + 517 z^5 - 160 z^6))  H(1, 0; y))/9 + (128 (-1 + y + z)^2 (66 y^8 + 198 y^7 (-1 + z) +  66 (-1 + z)^2 z^4 (1\nnn &
- z + z^2) + y^6 (264 - 588 z + 218 z^2) +  y^5 (-198 + 714 z - 628 z^2 + 163 z^3) +  6 y z^3 (14 - 68 z + 119 z^2 - 98 z^3 + 33 z^4)\nnn &
+  3 y^4 (22 - 136 z + 178 z^2 - 109 z^3 + 38 z^4) +  2 y^2 z^2 (-20 - 71 z + 267 z^2 - 314 z^3 + 109 z^4) +  y^3 z (84 - 142 z + 210 z^2\nnn &
- 327 z^3 + 163 z^4))  (H(1; z) H(3; y) + H(3, 2; y)))/9
\Big) \nnn
& + C_{2, g}\Big( (-128 (-1 + y) y (-1 + z)^2 (y + z)^2 (198 y^7 + y^6 (-792 + 1025 z) +  y^5 (1386 - 3992 z + 2187 z^2) + y^4 (-1386 + 6964 z\nnn &
- 8471 z^2 +  3016 z^3) + y^3 (792 - 6965 z + 14159 z^2 - 11110 z^3 + 3238 z^4) +  2 z (175 - 486 z + 797 z^2 - 836 z^3 + 525 z^4 \nnn & - 175 z^5)
+  y z (-1613 + 5147 z - 8266 z^2 + 6816 z^3 - 3126 z^4 + 692 z^5) +  y^2 (-198 + 4231 z - 12050 z^2 + 14766 z^3 - 8382 z^4\nnn &
+ 2076 z^5))  H(0; y))/27 - (128 (-1 + y)^2 (-1 + z) z (y + z)^2  (y^6 (-350 + 692 z) + 198 (-1 + z)^3 z^2 (1 - z + z^2) +  6 y^5 (175\nnn &
- 521 z + 346 z^2) + 2 y^3 (-1 + z)^2  (797 - 2539 z + 1508 z^2) + 2 y^4 (-836 + 3408 z - 4191 z^2 +  1619 z^3) + y (-1 + z)^2 (350 \nnn & - 913 z
+ 2055 z^2 - 1942 z^3 +  1025 z^4) + y^2 (-972 + 5147 z - 12050 z^2 + 14159 z^3 - 8471 z^4 +  2187 z^5)) H(0; z))/27\nnn & - (128 (-1 + y)^2 (-1
+ z)^2  (198 y^9 + y^8 (-792 + 757 z) + 198 (-1 + z)^3 z^4 (1 - z + z^2) +  y^7 (1386 - 3153 z + 1115 z^2)\nnn & + y^6 (-1386 + 5165 z - 4546 z^2
+  225 z^3) + y^5 (792 - 4466 z + 6493 z^2 - 1445 z^3 - 1383 z^4) +  y^4 (-198 + 1914 z - 5156 z^2 \nnn &+ 3014 z^3 + 1480 z^4 - 1383 z^5)
+  y^3 z (-217 + 1782 z - 4152 z^2  + 3014 z^3 - 1445 z^4 + 225 z^5) +  y z^3 (-217 + 1914 z - 4466 z^2 \nnn & + 5165 z^3 - 3153 z^4 + 757 z^5)
+  y^2 z^2 (-38 + 1782 z - 5156 z^2 + 6493 z^3 - 4546 z^4 + 1115 z^5))  (H(1; z) + H(2; y)))/27
\Big) \nnn
& + \Big( \frac{64 \, C_{1, g}}{27}  (2492 + 2306 y^4 + 4612 y^3 (-1 + z) - 4191 z + 6497 z^2 - 4612 z^3 +  2306 z^4 + y^2 (6497 - 13205 z + 6918 z^2)\nnn & +  y (-4191
+ 12784 z - 13205 z^2 + 4612 z^3))
\Big) \Bigg\}.\\
S_g^{(2), \, n_f^2}&=
  \bold{n_f^2}\, \Bigg\{  \frac{128 \, C_{1, g}}{9}  \Big((y^4 + 2 y^3 (-1 + z) + 3 y^2 (-1 + z)^2 + 2 y (-1 + z)^3 +  (1 - z + z^2)^2) (-3 \zeta_{2} - 4 H(0; y)^2 - 4 H(0; z)^2\nnn &
- 4 H(1; z)^2 -  4 H(0; y) (H(0; z) - H(1; z) - H(2; y)) - 8 H(1; z) H(2; y) -  4 H(2; y)^2 + 4 H(0; z) (H(1; z) + H(2; y))) \Big) \nnn
& +  \frac{256 \, C_{1, g}}{27} \Big((10 y^4 + y^3 (-20 + 23 z) + 3 y^2 (9 - 20 z + 11 z^2) +  y (-17 + 51 z - 57 z^2 + 20 z^3) + 2 (5 - 7 z + 12 z^2 - 10 z^3\nnn &
+  5 z^4)) H(0; y) +   (10 + 10 y^4 + 20 y^3 (-1 + z) - 17 z + 27 z^2 - 20 z^3 + 10 z^4 +  y (-1 + z)^2 (-14 + 23 z) + 3 y^2 (8
- 19 z \nnn & + 11 z^2)) H(0; z) -   (10 + 10 y^4 - 17 z + 27 z^2 - 20 z^3 + 10 z^4 + y^3 (-20 + 17 z) +  3 y^2 (9 - 17 z + 8 z^2) + y (-17
+ 54 z \nnn & - 51 z^2  + 17 z^3))  (H(1; z) + H(2; y))
\Big)\nnn
& -  \frac{128 \,C_{1, g}}{27}\Big( 3 y^4 + 6 y^3 (-1 + z) + y^2 (-7 + z + 9 z^2) +  z (10 - 7 z - 6 z^2 + 3 z^3) + y (10 - 17 z + z^2 + 6 z^3)
\Big)\Bigg\}. 
\end{align}
}
where the constants $C_{i,g}$ are
{\tiny 
\begin{align}
C_{1, g} &= \frac{Q^2}{y z (-1 + y + z)}\,, C_{2, g} =  \frac{Q^2} { y^2 z^2 (-1+y)^2 (-1+z)^2 (y+z)^2(-1+y+z)^2}\,, 
\nonumber\\
C_{3, g} &=  \frac{Q^2 }{  y^3 z^3  (-1+y+z)^3}\,,  C_{4, g} =  \frac{Q^2} { y^2 z^2 (-1+y) (-1+z) (y+z)(-1+y+z)^2}\,.
\end{align}
}
One loop finite result for $f=q\bar{q}g$ external state is given by
{\tiny
\begin{align}
 S_q^{G,(0)} &= -64\, C_{1, q}, \\
  S_q^{J,(0)} &= 0    ,\\
    S_q^{JG,(0)} &= 0  ,\\
 S_q^{JG,(1)} &=  - 64 \bold{n_f} \, C_{1, q}, \\
S_q^{G,(1)}&=
 \bold{N} \Bigg\{-128 \, C_{1, q} \big\{  H(0; y) H(1; z) + H(0; z) H(2; y) -  2 H(1; z) H(3; y) - H(0, 1; z) + H(0, 2; y) - H(1, 0; y) + H(2, 0; y) \nnn 
 & -  2 H(3, 2; y)
\big\}
 +  \frac{64 \, C_{1, q}} {3} \big\{10 H(0; y) + 10 H(0; z) - 13 H(1; z) - 13 H(2; y)\big\} 
   - \frac{64 \, C_{6, q}}{9} \big\{143 y^2 + y (9 - 18 z) + z (9 + 143 z) \big\}\Bigg\} 
\nnn &
+ \bold{n_f} \Bigg\{ \frac{-64 \, C_{1, q}}{3}\big\{  H(0; y) + H(0; z) - 4 H(1; z) - 4 H(2; y)\big\}
 +  \frac{1280 \, C_{1, q}}{9}\Bigg\}
 + \frac{1}{\bold{N}}\Bigg\{ -128 \, C_{1, q}   \big\{ \zeta_{2} + H(0; y) H(0; z) 
\nnn &
+ H(1, 0; y) + H(1, 0; z) \big\} 
 -64\, C_{6,q} (y + 7 y^2 + z - 2 y z + 7 z^2)\Bigg\}.
\end{align}.
}
At two loops, for $f=q\bar{q}g$ external state, we decompose ${S}_q^{G,(2)}$ in terms of the color factors
as follows
\begin{equation}
 S_q^{G,(2)} =  S_q^{(2), \, N^2} + S_q^{(2), \, N^0}+ S_q^{(2), \,1/ N^2}+ S_q^ {(2), \,n_f/N} + S_q^{(2), \, n_f  N}+  S_q^{(2), \, n_f^2}  \\
\end{equation} \\
where each color term is given below
{\tiny
\begin{align}
S_q^{(2), \, N^2} &=
  \bold{N^2}\Bigg\{ \frac{64 \, C_{1, q} }{5} \Big(  27 \zeta_{2}^2 - 80 \zeta_{3} H(2; y) +  40 \zeta_{2} H(0; z) H(2; y) - 10 H(0; y)^2 (H(0; z) + 4 H(1; z)) H(2; y)
+  10 \zeta_{2} ( H(0, 0; y)\nnn & +  H(0, 0; z) ) + 10 H(0, 0; y) H(0, 0; z)  -  20 \zeta_{2} H(0, 1; y) - 80 H(2; y) H(3; y) H(0, 1; z)
-  10 H(0, 0; y) H(0, 1; z) - 10 \zeta_{2} H(1, 0; y) \nnn &- 20 \zeta_{2} H(1, 1; y) +  70 \zeta_{2} H(1, 1; z) - 40 H(0, 0; y) H(1, 1; z)
-  10 H(0, 1; z) H(1, 2; y) + 10 \zeta_{2} H(2, 0; y) +  10 H(0, 0; z) H(2, 0; y)\nnn & - 10 H(0, 1; z) H(2, 0; y) +  10 \zeta_{2} H(2, 1; y)
+ 80 \zeta_{2} H(2, 2; y) - 40 H(0, 0; z) H(2, 2; y) -  80 H(0, 1; z) H(2, 2; y) + 20 \zeta_{2} H(3, 2; y)\nnn & -  20 H(0, 1; z) H(3, 2; y)
- 10 H(1, 0; y) H(3, 2; y) -  80 H(0, 1; z) H(3, 3; y) - 160 H(1, 1; z) H(3, 3; y) -  10 H(1; y) H(0, 0, 1; z) \nnn &- 40 H(2; y) H(0, 0, 1; z)
-  80 H(3; y) H(0, 0, 1; z) + 20 H(0; z) H(0, 0, 2; y) -  10 H(0; z) H(0, 1, 0; y) - 20 H(1; y) H(0, 1, 0; y)\nnn & +  30 H(2; y) H(0, 1, 0; z)
- 80 H(3; y) H(0, 1, 1; z) -  20 H(0; z) H(0, 2, 2; y) - 30 H(0; z) H(0, 3, 2; y) +  10 H(2; y) H(1, 0, 0; z)\nnn & - 10 H(1; y) H(1, 0, 1; z)
-  20 H(2; y) H(1, 0, 1; z) - 60 H(3; y) H(1, 0, 1; z) +  70 H(2; y) H(1, 1, 0; z) + 40 H(3; y) H(1, 1, 0; z)
\nnn &+  10 H(0; y) (3 \zeta_{2} H(2; y)  + H(2; y) H(0, 1; z) +  8 H(3; y) H(1, 1; z) + 4 H(1; z) (\zeta_{2} + H(2, 0; y) +  2 H(3, 2; y))
\nnn &+ H(0; z) (-(H(1; z) (2 H(2; y) + H(3; y))) +  H(2, 0; y)  + 2 (\zeta_{2} - H(2, 2; y) + H(3, 2; y))) + H(0, 0, 1; z) +  4 H(0, 1, 1; z)
\nnn &+ 3 H(0, 3, 2; y) + H(1, 0, 0; z) + 2 H(1, 0, 1; z) +  2 H(1, 1, 0; z)) + 10 H(0; z) H(2, 0, 0; y) +  20 H(0; z) H(2, 1, 0; y)
\nnn &+ 20 H(2; y) H(2, 1, 0; y) +  20 H(0; z) H(2, 2, 0; y) + 40 H(0; z) H(2, 3, 2; y) -  120 H(3; y) H(2, 3, 2; y) \nnn &+ 80 H(0; z) H(3, 2, 2; y)
-  10 H(1; z) (8 \zeta_{3} - 2 \zeta_{2} H(3; y) +  H(2; y) (-9 \zeta_{2}+ 12 H(3; y)^2) + H(3; y) H(1, 0; y)\nnn & -  12 H(3; y) H(3, 2; y)
- H(0; z) (3 \zeta_{2} + 4 H(2; y) H(3; y)  +  H(0, 0; y) - H(1, 0; y) + 4 H(2, 0; y) + 4 H(2, 2; y)\nnn & +  3 H(3, 0; y) + 2 H(3, 2; y))
- 4 H(0, 0, 2; y) - H(0, 1, 0; y) -  3 H(0, 2, 3; y) - 3 H(0, 3, 0; y)  - 4 H(1, 0, 0; y) + H(1, 2, 3; y) \nnn &+  H(1, 3, 0; y) + 4 H(2, 0, 0; y) 
- 4 H(2, 0, 3; y) - 2 H(2, 1, 0; y) +  4 H(2, 2, 3; y) - 4 H(2, 3, 0; y) - 16 H(2, 3, 3; y) +  H(3, 1, 0; y)\nnn & + 16 H(3, 3, 2; y)
+ 8 H(3, 3, 3; y)) -  50 H(0, 0, 0, 1; z) + 30 H(0, 0, 1, 0; z) - 40 H(0, 0, 1, 1; z) -  40 H(0, 0, 2, 2; y)\nnn & - 60 H(0, 0, 3, 2; y)
+ 20 H(0, 1, 0, 1; y) -  60 H(0, 1, 0, 1; z) + 10 H(0, 1, 0, 2; y) + 20 H(0, 1, 1, 0; y) +  20 H(0, 1, 1, 0; z)\nnn & + 10 H(0, 1, 2, 0; y)
- 40 H(0, 2, 0, 2; y) +  30 H(0, 2, 1, 0; y) - 40 H(0, 2, 2, 0; y) + 30 H(0, 2, 3, 2; y) +  80 H(0, 3, 2, 2; y) \nnn &- 30 H(1, 0, 0, 1; z)
+ 40 H(1, 0, 0, 2; y) +  30 H(1, 0, 1, 0; z) + 40 H(1, 0, 2, 0; y) - 40 H(1, 1, 0, 0; y) +  10 H(1, 1, 0, 0; z) \nnn &- 30 H(1, 1, 0, 1; z)
- 20 H(1, 1, 1, 0; y) +  40 H(1, 1, 1, 0; z) + 40 H(1, 2, 0, 0; y) - 10 H(1, 2, 1, 0; y) -  10 H(1, 2, 3, 2; y)\nnn & - 10 H(1, 3, 0, 2; y)
- 10 H(1, 3, 2, 0; y) -  40 H(2, 0, 0, 2; y) + 30 H(2, 0, 1, 0; y) - 40 H(2, 0, 2, 0; y) +  40 H(2, 0, 3, 2; y)\nnn & + 50 H(2, 1, 0, 0; y)
+ 10 H(2, 1, 1, 0; y) -  40 H(2, 2, 0, 0; y) - 40 H(2, 2, 3, 2; y) + 40 H(2, 3, 0, 2; y) +  40 H(2, 3, 2, 0; y) \nnn &+ 120 H(2, 3, 2, 3; y)
+ 160 H(2, 3, 3, 2; y) +  80 H(3, 0, 2, 2; y) - 10 H(3, 1, 0, 2; y) - 10 H(3, 1, 2, 0; y) +  80 H(3, 2, 0, 2; y)\nnn & - 10 H(3, 2, 1, 0; y)
+ 80 H(3, 2, 2, 0; y) -  160 H(3, 3, 2, 2; y) - 80 H(3, 3, 3, 2; y)
 \Big) \nnn
& + C_{4, q} \Big(-96 y z^3 H(0; y)^2 H(0; z) -  (64 y z (3 + 17 y^2 + 6 z - 4 z^2 - 6 y (1 + z)) \zeta_{2} H(1; z))/3 +  (32 y z (3 + 73 y^2 + 6 z
+ 61 z^2 \nnn & - 6 y (1 + z)) H(0; z) H(1; z)^2)/3 -  64 y z (2 + 7 y^2 - 4 y z + 7 z^2) \zeta_{2} H(2; y) +  (32 y z (3 + 53 y^2 - 6 y (-1 + z)\nnn &
- 6 z + 65 z^2) H(0; z)^2 H(2; y))/3 +  (128 y z (3 + 67 y^2 - 6 y z + 67 z^2) H(1; z) H(2; y) H(3; y))/3 -  (64 y z (3 + 61 y^2 + 6 z
+ 37 z^2 \nnn & - 6 y (1 + 2 z)) H(0; y) H(0, 1; z))/3 +  (64 y z (6 + 95 y^2 - 12 y z + 95 z^2) H(2; y) H(0, 1; z))/3 +  (64 y z (-3 + 5 y^2
+ 6 y (-1 + z) \nnn &+ 6 z  - 7 z^2) H(0; z) H(0, 2; y))/3 -  (64 y z (55 y^2 - 6 y (2 + z) + z (12 + 19 z)) H(1; z) H(0, 3; y))/3 +  (64 y z (-3
+ y^2 - 6 z + 13 z^2 \nnn & + 6 y (1 + z))  (H(1; z) H(1, 0; y) + H(2; y) H(1, 0; y)))/3 +  (128 y z (-3 + 7 y^2 + 6 y z
+ 7 z^2) H(1; z) H(3, 2; y))/3 -  (64 y z (43 y^2\nnn & - 6 y z  + 31 z^2) (H(1; z) H(3, 0; y) +  H(0; y) H(3, 2; y)))/3 - (64 y z (31 y^2 - 6 y z
+ 43 z^2)  (H(0; z) H(1; z) H(3; y) \nnn &+ H(0; z) H(3, 2; y)))/3  +  (64 y z (3 + 35 y^2 + 6 z - z^2 - 6 y (1 + 3 z)) H(0, 0, 1; z))/3
-  64 y z (1 - 2 y + 14 y^2 + 2 z \nnn &+ 17 z^2) H(0, 1, 0; y)  -  (64 (29 y^3 z  + 20 y z^3) H(0, 1, 0; z))/3 +  (64 y z (-3 + y^2 - 6 z + 13 z^2
+ 6 y (1 + z)) H(0, 1, 1; z))/3 \nnn & -  256 y z (-y + y^2 + z - z^2) H(0, 3, 2; y)  -  (64 y z (3 + 65 y^2 + 6 z + 62 z^2  - 6 y (1
+ z)) H(1, 0, 0; y))/3 \nnn & -  96 y^3 z (H(0; y) H(0; z)^2 + 2 H(1, 0, 0; z)) +  (64 y z (3 + 25 y^2 + 6 z + 4 z^2  - 6 y (1
+ z))  (\zeta_{2} H(1; y) + H(1, 1, 0; y)))/3 \nnn & -  (64 y z (3 + 91 y^2 + 6 z + 70 z^2 - 6 y (1 + z)) H(1, 1, 0; z))/3 +  (64 y z (3 + 65 y^2
+ 6 z + 53 z^2  - 6 y (1 + z))  ((H(0; y)^2 H(1; z))/2\nnn & + H(0; y) H(0, 2; y) - H(0, 0, 2; y) +  H(2, 0, 0; y)))/3 + (64 y z (3 + 38 y^2 + 6 z 
+ 26 z^2  - 6 y (1 + z))  H(2, 1, 0; y))/3 \nnn &+ (128 y z (3 + 67 y^2 - 6 y z + 67 z^2) H(2, 3, 2; y))/ 3 - 32 y z (y^2 + z^2) ((-638 \zeta_{3})/9
+ \zeta_{2} H(0; y)  + \zeta_{2} H(0; z) \nnn & -  (22 H(0; y) H(0; z) H(1; z))/3 + (74 H(0; y) H(1; z)^2)/3 +  (70 H(0; z) H(1; z) H(2; y))/3
+ (74 H(0; y) H(2; y)^2)/3 \nnn &+  (74 H(0; z) H(2; y)^2)/3 - (148 H(1; z)^2 H(3; y))/3 +  (148 H(1; z) H(0, 2; y))/3 + (40 H(0; z) H(1, 0; y))/3
\nnn & -  (62 H(0; z) H(2, 0; y))/3 + (148 H(1; z) H(2, 0; y))/3 -  (296 H(3, 2, 2; y))/3) \nnn &+ (512 y z (4 y^2 - 3 y z
+ 4 z^2)  ((H(1; z) H(3; y)^2)/2 + H(3; y) H(0, 1; z) + H(3, 3, 2; y)))/3
\Big) \nnn
& + C_{5, q}   \Big( (-32 y^2 z^2 (71 y^2 + 12 y (-1 + z) + z (-12 + 71 z)) \zeta_{2})/3 -  (32 y^2 z^2 (70 y^2 - 3 y z + z (3 + 67 z)) H(0; y)^2)/3
+  (32 y^2 z^2 (18 y  \nnn &- 263 y^2 + (18 - 263 z) z) H(0; y) H(0; z))/9 -  (32 y^2 z^2 (67 y^2 - 3 y (-1 + z) + 70 z^2) H(0; z)^2)/3
+  (32 y^2 z^2 (215 y^2 + 18 y (1 + z) \nnn & + z (18 + 197 z)) H(0; z) H(1; z))/9 +  (64 z^2 (-380 y^4 + 45 (-1 + z)^2 z^2 + 24 y^3 (1 + 3 z)
+  y^2 (48 - 291 z - 164 z^2) \nnn &+ 9 y z (11 - 16 z + 5 z^2)) H(0; z) H(2; y))/ 9 - (3328 y^2 z^2 (y^2 + z^2) (H(1; z)^2 + 2 H(1; z) H(2; y)
+ H(2; y)^2))/ 3 \nnn & - (32 y^2 (30 y^4 + 30 y^3 (-2 + z) + z^2 (32 + 28 z - 361 z^2) +  y^2 (30 - 96 z - 211 z^2) + 2 y z (33 - 91 z
+ 27 z^2)) H(0, 1; z))/3 \nnn &-  (32 y^2 (30 y^4 + 30 y^3 (-2 + z) + z^2 (32 + 10 z - 325 z^2) +  y^2 (30 - 96 z - 175 z^2) + 2 y z (33 - 100 z
+ 21 z^2)) H(1, 0; y))/3 \nnn & +  (64 y^2 (45 y^4 + 45 y^3 (-2 + z) + y^2 (45 - 144 z - 164 z^2) +  4 z^2 (12 + 6 z - 95 z^2) + 3 y z (33 - 97 z
+ 24 z^2))  (H(0; y) H(1; z)\nnn & + H(0, 2; y) + H(2, 0; y)))/9 -  (64 (15 y^6 + 15 y^5 (-2 + z) + 15 (-1 + z)^2 z^4 +  y^4 (15 - 48 z - 268 z^2)
+ 3 y z^3 (11 - 16 z + 5 z^2) \nnn & +  y^3 z (33 - 86 z + 48 z^2) - 2 y^2 z^2 (-16 + 43 z + 134 z^2))  (H(1; z) H(3; y) + H(3, 2; y)))/3
\Big) \nnn
& + {16\over27} \,C_{8, q} \Big(2 y (-1 + z) (270 y^5 + 2 (27 - 1813 z) z^3 +  y^4 (-540 + 4274 z) + y z^2 (216 - 4175 z \nnn &   + 3905 z^2)  +  y^2 z (432 - 4715 z
+ 4400 z^2) + y^3 (270 - 4706 z + 4499 z^2))  H(0; y) + (-1 + y) (2 z (270 (-1 + z)^2 z^3 + y^4 (-3626 \nnn &+ 3905 z)  +  2 y z^2 (216 - 2353 z
+ 2137 z^2) + y^3 (54  - 4175 z + 4400 z^2) +  y^2 z (216 - 4715 z + 4499 z^2)) H(0; z) \nnn &+  3 (-1 + z) (180 y^5  + 180 (-1 + z)^2 z^3 - y^4 (360
+ 4357 z) +  y z^2 (852 - 966 z  - 4357 z^2) + y^2 z (852 - 3420 z - 2821 z^2) \nnn & +  y^3 (180 - 966 z - 2821 z^2)) (H(1; z)
+ H(2; y)))\Big) \nnn
& -{4\over81}C_{6, q} \Big( 648 + 393245 y^2 + 52272 z + 393245 z^2 - 432 y (-121 + 245 z)\Big)
\Bigg\}.
\end{align}
}
{\tiny
\begin{align}
S_q^{(2), \, N^0} &=
  \Bigg\{  -32 \, C_{1, q} \Big(7 \zeta_{2}^2 - 12 \zeta_{3} H(1; y) - 16 \zeta_{3} H(1; z) +  4 \zeta_{2} H(1; y) H(1; z) - 4 \zeta_{3} H(2; y)
+ 4 H(0; y)^2 H(0; z)  H(2; y) \nnn & - 8 \zeta_{2} H(1; z) H(3; y) + 8 H(1; z) H(3; y) H(0, 0; y) -  16 H(0, 0; y) H(0, 0; z)
- 16 \zeta_{2} H(0, 1; y) - 4 \zeta_{2} H(0, 1; z) -  16 \zeta_{2} H(1, 0; y) \nnn & - 4 H(1; z) H(3; y) H(1, 0; y) -  16 \zeta_{2} H(1, 1; y)
- 12 \zeta_{2} H(1, 1; z) + 4 \zeta_{2} H(1, 2; y) -  8 H(0, 1; z) H(1, 2; y) - 4 \zeta_{2} H(2, 0; y) \nnn & +  4 H(0, 1; z) H(2, 0; y)
- 8 \zeta_{2} H(3, 2; y) +  8 H(0, 1; z) H(3, 2; y) - 4 H(1, 0; y) H(3, 2; y) -  4 H(1; y) H(0, 0, 1; z)\nnn & - 24 H(1; y) H(0, 1, 0; y)
+  8 H(1; z) H(0, 1, 0; y) + 4 H(1; y) H(0, 1, 0; z) +  4 H(2; y) H(0, 1, 0; z) - 4 H(1; z) H(0, 2, 3; y)\nnn & -  4 H(1; z) H(0, 3, 0; y)
+ 8 H(1; z) H(1, 0, 0; y) +  8 H(2; y) H(1, 0, 0; z) - 4 H(1; y) H(1, 0, 1; z) -  8 H(3; y) H(1, 0, 1; z)\nnn & - 4 H(0; y) (3 \zeta_{3}
- 2 \zeta_{2} H(1; z) -  2 \zeta_{2} H(2; y) - 2 H(2; y) H(0, 0; z) + 2 H(2; y) H(0, 1; z) +  H(0; z) (-(H(1; z) (H(2; y)\nnn & - 3 H(3; y)))
 + H(2, 0; y) +  2 (\zeta_{2} + H(3, 2; y))) + 5 H(0, 0, 1; z) + 3 H(0, 1, 0; z) -  H(0, 3, 2; y) + 4 H(1, 0, 0; z) + H(1, 0, 1; z)
\nnn &- H(1, 1, 0; z)) +  4 H(1; y) H(1, 1, 0; z)  - 16 H(3; y) H(1, 1, 0; z) -  4 H(1; z) H(1, 2, 3; y) - 4 H(1; z) H(1, 3, 0; y)
-  4 H(0; z) (3 \zeta_{3}\nnn & - \zeta_{2} H(1; y) - \zeta_{2} H(2; y) +  H(2; y) H(1, 0; y) + H(1; z) (3 \zeta_{2} - 2 H(0, 0; y) + 2 H(1, 0; y)
-  H(1, 2; y) - H(3, 0; y) + 2 H(3, 2; y))\nnn & + 2 H(0, 0, 2; y) +  2 H(0, 1, 0; y) + H(0, 3, 2; y) + 6 H(1, 0, 0; y) - H(1, 2, 0; y)
-  2 H(2, 0, 0; y) - 2 H(2, 1, 0; y))\nnn & - 8 H(1; z) H(3, 0, 0; y)-  4 H(1; z) H(3, 1, 0; y)  + 12 H(0, 0, 0, 1; z) - 8 H(0, 0, 1, 0; y)
-  4 H(0, 0, 1, 0; z) + 24 H(0, 1, 0, 1; y)\nnn & - 4 H(0, 1, 0, 1; z) +  8 H(0, 1, 0, 2; y) + 32 H(0, 1, 1, 0; y) - 16 H(0, 1, 1, 0; z)
+  8 H(0, 1, 2, 0; y) + 8 H(0, 2, 1, 0; y)\nnn & - 4 H(0, 2, 3, 2; y) -  4 H(1, 0, 0, 1; z) + 8 H(1, 0, 0, 2; y) - 20 H(1, 0, 1, 0; z)
+  8 H(1, 0, 2, 0; y)  - 32 H(1, 1, 0, 0; y)\nnn & - 16 H(1, 1, 0, 0; z) -  4 H(1, 1, 0, 1; z) - 16 H(1, 1, 1, 0; y) - 16 H(1, 1, 1, 0; z)
+  8 H(1, 2, 0, 0; y) - 4 H(1, 2, 3, 2; y)\nnn & - 4 H(1, 3, 0, 2; y) -  4 H(1, 3, 2, 0; y) + 4 H(2, 0, 1, 0; y) + 8 H(2, 1, 0, 0; y)
-  4 H(3, 1, 0, 2; y) - 4 H(3, 1, 2, 0; y)\nnn & - 4 H(3, 2, 1, 0; y)\Big)
  \nnn
 & + C_{2, q} \Big((32 (761 (-1 + z)^2 z^2 - 2 y z^2 (869 - 1630 z + 761 z^2) +  y^4 (761 - 1522 z + 815 z^2) + 2 y^3 (-761 + 1630 z - 1139 z^2\nnn &
+  216 z^3) + y^2 (761 - 1738 z + 2494 z^2 - 2278 z^3 + 815 z^4)) \zeta_{3})/ 9 + (64 (-1 + y)^2 (-12 y (-1 + z) z^2 + 11 (-1 + z)^2 z^2\nnn &
+  y^2 (38 - 76 z + 35 z^2)) \zeta_{2} H(0; y))/3 +  (64 (-1 + z)^2 (11 y^4 + 38 z^2 - 76 y z^2 - 2 y^3 (11 + 6 z) +  y^2 (11 + 12 z\nnn &
+ 35 z^2)) \zeta_{2} H(0; z))/3 +  (64 (-1 + y)^2 (-1 + z)^2 (31 y^2 + 22 z^2) H(0; y)^2 H(0; z))/3 -  (32 ((-1 + z)^2 z (-36 + 103 z)\nnn &
+ y^4 (31 - 62 z + 37 z^2) +  2 y z (54 - 223 z + 272 z^2 - 103 z^3) +  2 y^3 (-31 + 92 z - 109 z^2 + 42 z^3) +  y^2 (31 - 194 z + 458 z^2\nnn &
- 398 z^3 + 109 z^4)) \zeta_{2} H(1; z))/3 +  (128 (-1 + y)^2 (-1 + z)^2 (8 y^2 - z^2) H(0; y) H(0; z) H(1; z))/3 -  (32 (13 (-1 + z)^2 z^2\nnn &
+ y^4 (13 - 26 z + 10 z^2) -  2 y z^2 (7 - 20 z + 13 z^2) + y^3 (-26 + 40 z + 16 z^2 - 24 z^3) +  y^2 (13 - 14 z - 28 z^2 + 16 z^3\nnn &
+ 10 z^4)) H(0; z) H(1; z)^2)/3 -  (32 ((-1 + z)^2 z (-36 + 59 z) + y^4 (59 - 118 z + 65 z^2) +  2 y^3 (-77 + 202 z - 191 z^2 + 60 z^3)\nnn &
-  2 y (18 - 108 z + 233 z^2 - 202 z^3 + 59 z^4) +  y^2 (131 - 466 z + 658 z^2 - 382 z^3 + 65 z^4)) \zeta_{2} H(2; y))/3 -  192 (-1\nnn &
+ y)^2 y (-1 + z)^2 (-1 + y + z) H(0; z)^2 H(2; y) -  (64 (-1 + z)^2 (31 y^4 + 4 z^2 + 10 y^3 (-8 + 3 z) -  2 y (9 - 9 z + 4 z^2) + y^2 (67\nnn &
- 48 z + 7 z^2)) H(0; z) H(1; z)  H(2; y))/3 - (64 (-1 + y)^2 (2 (-1 + z)^2 z (9 + z) -  6 y z (1 - 4 z + 3 z^2) + y^2 (41 - 82 z\nnn &
+ 38 z^2)) H(0; y) H(0, 1; z))/ 3 + (64 (13 (-1 + z)^2 z^2 + y^4 (40 - 80 z + 43 z^2) +  y^3 (-98 + 238 z - 200 z^2 + 54 z^3) -  2 y (9\nnn &
- 27 z + 46 z^2 - 41 z^3 + 13 z^4) +  y^2 (76 - 212 z + 242 z^2 - 122 z^3 + 19 z^4)) H(2; y) H(0, 1; z))/3 +  (64 (-1 + z)^2 (5 y^4 - 22 z^2\nnn &
+ y^3 (-28 + 30 z) +  y^2 (41 - 48 z - 19 z^2) + 2 y (-9 + 9 z + 22 z^2)) H(0; z) H(0, 2; y))/ 3 - 64 (-6 (-1 + z)^3 z + y^4 (10 - 20 z\nnn &
+ 9 z^2) +  y^3 (-26 + 60 z - 36 z^2 + 4 z^3) +  y^2 (22 - 50 z + 20 z^2 + 12 z^3 - 5 z^4) +  2 y (-3 + 2 z + 13 z^2 - 18 z^3\nnn &
+ 6 z^4)) H(1; z) H(0, 3; y) +  (128 (-1 + y)^2 (-1 + z)^2 (y^2 + 10 z^2) H(0; z) H(1, 0; y))/3 -  (1408 (-1 + y)^2 (-1 + z)^2 (y^2\nnn &
+ z^2) (H(1; z) H(1, 0; y) +  H(2; y) H(1, 0; y) + (2 H(0; z) H(2, 0; y))/11))/3 -  64 (-1 + y)^2 (-1 + z)^2 (y^2 + 4 y z\nnn &
+ 3 z^2)  (H(1; z) H(3, 0; y) + H(0; y) H(3, 2; y)) - 64 (-1 + y)^2 (-1 + z)^2  (3 y^2 + 4 y z + z^2) (H(0; z) H(1; z) H(3; y)\nnn &
+ H(0; z) H(3, 2; y)) +  64 (-3 (-2 + z) (-1 + z)^2 z + y^4 (5 - 10 z + 4 z^2) +  2 y^3 (-5 + 13 z - 8 z^2 + z^3) + 2 y z (-5 + 15 z - 13 z^2\nnn &
+ 3 z^3) +  y^2 (5 - 12 z - 2 z^2 + 10 z^3 - 2 z^4)) H(0, 0, 1; z) +  64 (-1 + y)^2 (2 (-1 + z)^2 z (-3 + 10 z) + 2 y z (5 - 8 z + 3 z^2)\nnn &
+  y^2 (15 - 30 z + 14 z^2)) H(0, 1, 0; y) +  64 (-1 + z)^2 (11 y^4 + 14 z^2 - 28 y z^2 - 2 y^3 (11 + 2 z) +  y^2 (11 + 4 z\nnn &
+ 13 z^2)) H(0, 1, 0; z) +  (64 (13 (-1 + z)^2 z^2 + y^4 (13 - 26 z + 10 z^2) -  2 y z^2 (7 - 20 z + 13 z^2) + y^3 (-26 + 40 z + 16 z^2\nnn &
- 24 z^3) +  y^2 (13 - 14 z - 28 z^2 + 16 z^3 + 10 z^4)) H(0, 1, 1; z))/3 -  64 (-3 (-1 + z)^2 z (-2 + 3 z) + y^3 (-24 + 52 z - 26 z^2)\nnn &
+  y^4 (9 - 18 z + 8 z^2) + y^2 (21 - 40 z + 26 z^3 - 8 z^4) +  2 y (-3 + 20 z^2 - 26 z^3 + 9 z^4)) H(0, 3, 2; y) +  (128 (-1 + y)^2 (-1\nnn &
+ z)^2 (31 y^2 + 9 y z + z (-9 + 31 z)) H(1, 0, 0; y))/ 3 + (64 (-1 + y)^2 (-1 + z)^2 (22 y^2 + 31 z^2)  (H(0; y) H(0; z)^2\nnn &
+ 2 H(1, 0, 0; z)))/3 -  (128 (-1 + y)^2 (-1 + z)^2 (11 y^2 - 9 y z + (9 - 7 z) z)  (\zeta_{2} H(1; y) + H(1, 1, 0; y)))/3 -  (64 (35 (-1\nnn &
+ z)^2 z^2 + y^4 (26 - 52 z + 29 z^2) -  2 y z^2 (41 - 76 z + 35 z^2) + 2 y^3 (-26 + 64 z - 59 z^2 + 18 z^3) +  y^2 (26 - 76 z + 142 z^2\nnn &
- 130 z^3 + 41 z^4)) H(1, 1, 0; z))/3 -  384 (-1 + y)^2 (-1 + z)^2 z (-1 + y + z) ((H(0; y)^2 H(1; z))/2 +  H(0; y) H(0, 2; y)\nnn &
- H(0, 0, 2; y) + H(2, 0, 0; y)) +  64 (-1 + y)^2 (-3 (-2 + z) (-1 + z)^2 z + y^2 (6 - 12 z + 5 z^2) -  2 y z (3 - 8 z\nnn &
+ 5 z^2)) H(2, 1, 0; y) +  64 (3 (-1 + z)^2 z^2 - 2 y z^2 (5 - 8 z + 3 z^2) + y^4 (3 - 6 z + 4 z^2) +  2 y^3 (-3 + 8 z - 10 z^2 + 4 z^3)\nnn &
+  y^2 (3 - 10 z + 24 z^2 - 20 z^3 + 4 z^4)) (H(1; z) H(2; y) H(3; y) -  H(1; z) H(3, 2; y) + H(2, 3, 2; y)) + 512 (-1 + y)^2 (-1 + z)^2 (y\nnn &
+ z)^2  ((H(1; z) H(3; y)^2)/2 + H(3; y) H(0, 1; z) + H(3, 3, 2; y))
\Big) \nnn
 & + C_{3, q} \Big( -64 y^2 z^2 (43 y^4 (-1 + z) - 43 (-1 + z)^2 z^2 +  y z^2 (90 - 133 z + 43 z^2) + y^3 (86 - 133 z + 48 z^2) +  y^2 (-43 + 90 z - 96 z^2\nnn &
+ 48 z^3)) \zeta_{2} -  (32 (-1 + y) y^2 (-1 + z) z^2 (307 y^3 + 325 y z^2 + 307 (-1 + z) z^2 +  y^2 (-307 + 325 z)) H(0; y) H(0; z))/3\nnn &
-  (32 y^2 (-1 + z) z^2 (307 y^4 - 307 (-1 + z) z^2 + y^3 (-614 + 325 z) +  y z (-6 - 626 z + 307 z^2) + y^2 (307 - 319 z\nnn &
+ 325 z^2)) H(0; z)  H(1; z))/3 - 32 (-1 + z) z^2 (33 y^6 + (-1 + z)^3 z^2 +  y^5 (-64 + 39 z) - y (-1 + z)^2 z (2 - 10 z + z^2) +  y^4 (26\nnn &
- 41 z + 36 z^2) + y^3 (8 - 5 z - 52 z^2 + 22 z^3) +  y^2 (-3 + 9 z + 3 z^2 - 9 z^4)) H(0; z) H(2; y) -  (32 y^2 (3 y^6 (-1 + z) + 3 y^5 (4
\nnn &- 13 z + 9 z^2)  +  (-1 + z)^2 z^2 (-9 + 6 z + 406 z^2) +  y^4 (-18 + 105 z + 280 z^2 - 373 z^3) +  y^3 (12 - 111 z - 623 z^2 + 1155 z^3\nnn &
- 433 z^4)  +  y^2 (-3 + 48 z  + 289 z^2 - 749 z^3 + 863 z^4 - 442 z^5) -  y z (6 - 36 z + 57 z^2 + 803 z^3 - 1236 z^4
+ 406 z^5)) H(0, 1; z))/3 \nnn &-  (32 y^2 (3 y^6 (-1 + z)  + 3 y^5 (4 - 13 z + 9 z^2) -  (-1 + z)^2 z^2 (9 - 6 z + 208 z^2) +  y^4 (-18 + 105 z
- 334 z^2 + 241 z^3) +  y^3 (12 \nnn &- 111 z + 605 z^2 - 711 z^3 + 217 z^4) +  y z (-6 + 36 z - 33 z^2 + 425 z^3 - 630 z^4 + 208 z^5) +  y^2 (-3
+ 48 z - 325 z^2 + 479 z^3 \nnn &- 413 z^4 + 208 z^5)) H(1, 0; y))/3  +  32 y^2 (y^6 (-1 + z) + y^5 (4 - 13 z + 9 z^2) +  (-1 + z)^2 z^2 (-3 + 2 z
+ 33 z^2)\nnn & - y^4 (6 - 35 z + 9 z^2 + 22 z^3) +  y^3 (4 - 37 z - 3 z^2 + 74 z^3  - 36 z^4) -  y z (2 - 12 z + 13 z^2 + 67 z^3 - 103 z^4
+ 33 z^5)\nnn & -  y^2 (1 - 16 z + 6 z^2 + 47 z^3 - 77 z^4 + 39 z^5))  (H(0; y) H(1; z) + H(0, 2; y) + H(2, 0; y)) -  32 (y^8 (-1 + z)\nnn & - (-1
+ z)^4 z^4 + y (-1 + z)^3 z^3 (2 - 10 z + z^2) +  y^7 (4 - 13 z + 9 z^2) + y^6 (-6 + 35 z + 24 z^2 - 55 z^3) +  y^2 (-1 + z)^2 z^2 (-6\nnn & + 8 z
+ 42 z^2 + 9 z^3) +  y^5 (4 - 37 z - 67 z^2 + 175 z^3 - 75 z^4) +  y^4 (-1 + 16 z + 20 z^2 - 108 z^3 + 150 z^4 - 75 z^5)\nnn & -  y^3 z (2 - 20 z
+ 30 z^2 + 108 z^3 - 175 z^4 + 55 z^5))  (H(1; z) H(3; y) + H(3, 2; y))
\Big) \nnn
 & + {32\over27}\,C_{7, q}\Big(  (9 y (-1 + z) (3 y^4 + y^3 (-6 + 145 z) - z^2 (11 + 149 z) +  y^2 (3 - 109 z - 69 z^2) 
 \nnn &+ y z (-36 + 80 z + 109 z^2)) H(0; y)
+  (-1 + y) (9 z (3 (-1 + z)^2 z^2 + y^3 (-149 + 109 z) +  y^2 (-11 + 80 z - 69 z^2) + y z (-36 \nnn &- 109 z + 145 z^2)) H(0; z) +  (-1
+ z) (27 y^4 + 27 (-1 + z)^2 z^2 - 2 y^3 (27 + 1547 z) +  9 y^2 (3 - 46 z + 38 z^2) - 2 y z (-108 + 207 z \nnn &+ 1547 z^2))  (H(1; z)
+ H(2; y))))
\Big) \nnn
&  -{16\over81}C_{6, q}  \Big( 324 + 56327 y^2 + 14040 z + 56327 z^2 - 216 y (-65 + 133 z)\Big)
 \Bigg\}.
\end{align}
}
{\tiny
\begin{align}
S_q^{(2), \, 1/N^2} &=
\bold{\frac{1}{N^2}}\Bigg\{ \frac{-128 \,C_{1, q}}{5} \Big( 37 \zeta_{2}^2 - 15 \zeta_{3} H(1; y) - 55 \zeta_{3} H(1; z) +  5 \zeta_{2} H(1; y) H(1; z) - 40 \zeta_{3} H(2; y)
+ 5 \zeta_{2} H(1; z) H(2; y) \nnn & +  5 \zeta_{2} H(0, 0; y)  - 10 H(1; z) H(3; y) H(0, 0; y) +  5 \zeta_{2} H(0, 0; z) + 5 H(0, 0; y) H(0, 0; z)
+ 10 \zeta_{2} H(0, 1; y) -  5 \zeta_{2} H(0, 1; z)\nnn & - 5 H(0, 0; y) H(0, 1; z) + 15 \zeta_{2} H(1, 0; y) +  10 H(1; z) H(3; y) H(1, 0; y)
+ 10 \zeta_{2} H(1, 1; y) +  20 \zeta_{2} H(1, 1; z) + 5 \zeta_{2} H(1, 2; y)\nnn & - 15 H(0, 1; z) H(1, 2; y) +  5 H(0, 0; z) H(2, 0; y)
+ 5 \zeta_{2} H(2, 1; y) +  10 H(1, 0; y) H(3, 2; y) - 10 H(1; y) H(0, 0, 1; z)\nnn & +  20 H(1; y) H(0, 1, 0; y) - 5 H(1; z) H(0, 1, 0; y)
+  5 H(1; y) H(0, 1, 0; z) - 10 H(1; z) H(0, 2, 3; y) +  10 H(1; z) H(0, 3, 0; y)\nnn & - 10 H(1; z) H(1, 0, 0; y) -  5 H(2; y) H(1, 0, 0; z)
- 10 H(1; y) H(1, 0, 1; z) -  10 H(2; y) H(1, 0, 1; z) + 5 H(1; y) H(1, 1, 0; z) \nnn & -  5 H(2; y) H(1, 1, 0; z) - 5 H(0; y) (4 \zeta_{3}
- 2 \zeta_{2} H(1; z) -  \zeta_{2} H(2; y) + 2 H(2; y) H(0, 0; z) + H(2; y) H(0, 1; z) +  H(0; z) (-4 \zeta_{2}\nnn & + H(1; z) H(2; y)
+ 2 H(2, 2; y)) - H(0, 1, 0; z) -  H(1, 0, 0; z) + 3 H(1, 0, 1; z) + H(1, 1, 0; z)) -  10 H(1; z) H(1, 2, 3; y)\nnn & - 10 H(1; z) H(1, 3, 0; y)
-  10 H(1; z) H(2, 1, 0; y) - 10 H(2; y) H(2, 1, 0; y) -  5 H(0; z) (4 \zeta_{3} - \zeta_{2} H(1; y) \nnn & - \zeta_{2} H(2; y)
+  H(2; y) H(1, 0; y) - H(1; z) (4 \zeta_{2} - H(0, 0; y) + H(1, 0; y)  +  H(1, 2; y)) - H(0, 1, 0; y) - 2 H(0, 2, 2; y)\nnn &  - 2 H(1, 0, 0; y)
-  H(1, 2, 0; y) + H(2, 0, 0; y) - 2 H(2, 2, 0; y)) +  10 H(1; z) H(3, 0, 0; y)  - 10 H(1; z) H(3, 1, 0; y)\nnn & -  10 H(0, 0, 0, 1; z) 
+ 10 H(0, 0, 1, 0; y) + 10 H(0, 0, 1, 0; z) -  10 H(0, 0, 3, 2; y) - 20 H(0, 1, 0, 1; y) - 15 H(0, 1, 0, 1; z)\nnn & -  5 H(0, 1, 0, 2; y)
- 30 H(0, 1, 1, 0; y) + 10 H(0, 1, 1, 0; z) -  5 H(0, 1, 2, 0; y) + 5 H(0, 2, 1, 0; y) - 10 H(0, 2, 3, 2; y) \nnn & -  10 H(1, 0, 0, 2; y)
+ 10 H(1, 0, 1, 0; z) - 10 H(1, 0, 2, 0; y) +  20 H(1, 1, 0, 0; y) + 5 H(1, 1, 0, 0; z) - 20 H(1, 1, 0, 1; z) \nnn &+  10 H(1, 1, 1, 0; y)
- 10 H(1, 2, 0, 0; y) - 5 H(1, 2, 1, 0; y) -  10 H(1, 2, 3, 2; y) - 10 H(1, 3, 0, 2; y) - 10 H(1, 3, 2, 0; y) \nnn & -  5 H(2, 1, 0, 0; y)
+ 5 H(2, 1, 1, 0; y) + 20 H(2, 2, 1, 0; y) -  10 H(3, 1, 0, 2; y) - 10 H(3, 1, 2, 0; y) - 10 H(3, 2, 1, 0; y)
\Big) \nnn
& + C_{2, q} \Big(192 (9 (-1 + z)^2 z^2 - 2 y z^2 (11 - 20 z + 9 z^2) +  y^4 (9 - 18 z + 10 z^2) + 2 y^3 (-9 + 20 z - 16 z^2 + 4 z^3) +  y^2 (9 - 22 z\nnn &
+ 36 z^2 - 32 z^3 + 10 z^4)) \zeta_{3} -  64 (-1 + y)^2 (4 y (-1 + z) z^2 + 6 (-1 + z)^2 z^2 + y^2 (3 - 6 z + 4 z^2))  \zeta_{2} H(0; y)\nnn &
- 96 (-1 + y)^2 (-1 + z)^2 z^2 H(0; y)^2 H(0; z) -  64 ((-1 + z)^2 (-1 + 4 z + 9 z^2) + y^4 (14 - 28 z + 15 z^2) +  y^3 (-26 + 52 z \nnn & - 32 z^2
+ 4 z^3) +  y (4 - 20 z + 6 z^2 + 28 z^3 - 18 z^4) +  y^2 (9 - 10 z + 12 z^2 - 20 z^3 + 10 z^4)) \zeta_{2} H(1; z) +  32 ((-1 + z)^2 (-1
+ 4 z)\nnn & + y^4 (2 - 4 z + 3 z^2) -  4 y (-1 + 5 z - 6 z^2 + 2 z^3) + y^3 (-2 + 4 z - 8 z^2 + 4 z^3) +  y^2 (-3 + 14 z - 9 z^2 \nnn & - 2 z^3
+ z^4)) H(0; z) H(1; z)^2  -  64 (-2 y^3 (5 - 8 z + 4 z^2) + 2 (-1 + z)^2 (-1 + 3 z + 4 z^2) +  y^4 (8 - 16 z + 9 z^2) \nnn & - 2 y (-5 + 20 z
- 15 z^2 - 8 z^3 + 8 z^4) +  y^2 (-6 + 30 z  - 24 z^2  - 8 z^3 + 9 z^4)) \zeta_{2} H(2; y) -  32 (-1 + y)^2 (3 y^2 + 4 y (-1 + z) \nnn & + (-1
+ z)^2) (-1 + z)^2 H(0; z)^2  H(2; y)  - 64 (-1 + z)^2 (6 y^4  + 4 y^3 (-3 + z) + 3 z^2\nnn & - 6 y z^2 +  y^2 (6 - 4 z + 4 z^2)) (\zeta_{2} H(0; z)
+ H(0; z) H(1; z) H(2; y)) +  64 (-1 + y)^2 (y^2 (4 - 8 z + 5 z^2)  + (-1 + z)^2 (1 - 4 z \nnn &+ 9 z^2) +  2 y (-1 + 4 z - 7 z^2
+ 4 z^3)) H(0; y) H(0, 1; z) +  64 (2 (-1 + z)^2 (-1 + 3 z + z^2) + y^4 (5 - 10 z + 6 z^2) \nnn & +  2 y^3 (-2 + 4 z - 5 z^2 + 2 z^3) - 2 y (-5
+ 20 z - 21 z^2 + 4 z^3 +  2 z^4) + y^2 (-9 + 32 z - 24 z^2 - 2 z^3 + 4 z^4)) H(2; y) H(0, 1; z)\nnn & -  64 (-1 + y)^2 (-1 + z)^2 (1 + y^2 - 4 z
+ 3 z^2 + y (-2 + 4 z)) H(0; y)  H(0, 2; y)  + 64 (-1 + z)^2 (6 y^4 + 8 y^3 (-2 + z) + (-1 + z)^2\nnn & -  2 y (3 - 4 z + z^2) + y^2 (15 - 14 z
+ 2 z^2)) H(0; z) H(0, 2; y) -  192 (-1 + y)^2 (-1 + z)^2 (y^2 + z^2) (H(0; y) H(0; z) H(1; z)\nnn & +  H(0; z) H(2, 0; y)) + 64 (y^4 (2 - 4 z
+ z^2) -  (-1 + z)^2 (1 - 4 z + 6 z^2) + y^3 (-2 + 4 z + 4 z^2 - 4 z^3) +  y^2 (-3 + 14 z - 33 z^2 \nnn & + 26 z^3  - 5 z^4)  +  4 y (1 - 5 z + 11 z^2 
- 10 z^3 + 3 z^4)) H(0, 0, 1; z) -  32 (-1 + y)^2 (-1 + z)^2 (1 + y^2 - 4 z + 3 z^2 \nnn &+ y (-2 + 4 z))  (H(0; y)^2 H(1; z)
- 2 H(1; z) H(1, 0; y) - 2 H(2; y) H(1, 0; y) -  2 H(0, 0, 2; y)) - 64 (-1 + y)^2 (y^2 (-1 + 2 z) \nnn &+  (-1 + z)^2 (-1 + 4 z + 3 z^2) + y (2
- 8 z + 6 z^2)) H(0, 1, 0; y) -  64 y^2 (-1 + z)^2 (6 + 6 y^2 + 4 y (-3 + z) - 4 z  \nnn &+ z^2) H(0, 1, 0; z) -  64 ((-1 + z)^2 (-1 + 4 z) + y^4 (2
- 4 z + 3 z^2) -  4 y (-1 + 5 z - 6 z^2 + 2 z^3) + y^3 (-2 + 4 z - 8 z^2 + 4 z^3)\nnn & +  y^2 (-3 + 14 z - 9 z^2 - 2 z^3 + z^4)) H(0, 1, 1; z)
-  64 (-((-1 + z)^2 z (-2 + 5 z)) + y^4 (5 - 10 z + 4 z^2) -  2 y^3 (6 - 14 z + 7 z^2)\nnn &  + y^2 (9 - 20 z + 14 z^3 - 4 z^4) +  2 y (-1 + 10 z^2
- 14 z^3 + 5 z^4)) (H(1; z) H(0, 3; y) +  H(0, 3, 2; y))\nnn & + 64 (-1 + y)^3 (-1 + z)^2 (-1 + y + 4 z) H(1, 0, 0; y) -  96 (-1 + y)^2 y^2 (-1 
+ z)^2 (H(0; y) H(0; z)^2 \nnn & + 2 H(1, 0, 0; z)) -  64 (-1 + y)^2 (-1 + z)^2 (1 + y^2 - 4 z + 6 z^2 + y (-2 + 4 z))  (\zeta_{2} H(1; y) \nnn &
+ H(1, 1, 0; y)) -  64 ((-1 + z)^2 (-1 + 4 z + 3 z^2) + y^4 (11 - 22 z + 12 z^2) +  y^3 (-20 + 44 z - 34 z^2 \nnn &+ 8 z^3) +  y (4 - 20 z + 18 z^2
+ 4 z^3 - 6 z^4) +  y^2 (6 - 8 z + 12 z^2 - 14 z^3 + 5 z^4)) H(1, 1, 0; z) -  64 (-1 + y)^2 (-1 + z)^2 (1 + y^2 \nnn &- 4 z + 3 z^2  + y (-2
+ 4 z))  H(2, 0, 0; y) - 64 (-1 + y)^2 (y^2 (4 - 8 z + 5 z^2) +  (-1 + z)^2 (1 - 4 z + 9 z^2) + 2 y (-1 + 4 z - 7 z^2
\nnn &+ 4 z^3))  H(2, 1, 0; y) + 64 (y^4 (-1 + 2 z) - (-1 + z)^2 (2 - 6 z + z^2) +  2 y^3 (4 - 10 z + 5 z^2) + y^2 (-15 + 48 z - 42 z^2 + 10 z^3)
\nnn & +  2 y (5 - 20 z + 24 z^2  - 10 z^3 + z^4)) (H(1; z) H(2; y) H(3; y) -  H(1; z) H(3, 2; y) + H(2, 3, 2; y))
\Big) \nnn
& + C_{3, q} \Big( -32 y^2 z^2 (55 y^4 (-1 + z) - (-1 + z)^2 z (4 + 55 z) +  y^3 (106 - 153 z + 49 z^2) + y^2 (-47 + 82 z - 86 z^2 + 49 z^3) +  y (-4\nnn &
+ 20 z + 82 z^2 - 153 z^3 + 55 z^4)) \zeta_{2} -  32 (-1 + y) y^2 (-1 + z) z^3 (-1 + y + z)^2 H(0; y)^2 -  64 (-1 + y) y^2 (-1
+ z) z^2 (13 y^3 \nnn &+ 2 y^2 (-6 + 5 z) +  y (-1 + 2 z + 10 z^2) + z (-1 - 12 z + 13 z^2)) H(0; y) H(0; z)\nnn & -  32 (-1 + y) y^3 (-1 + z) z^2 (-1
+ y + z)^2 H(0; z)^2 -  64 y^2 (-1 + z) z^2 (14 y^4 + z + 12 z^2 - 13 z^3 \nnn &+ 3 y^3 (-9 + 4 z) +  y^2 (12 - 10 z + 11 z^2) + y (1 - 3 z
- 22 z^2 + 13 z^3)) H(0; z)  H(1; z)  - 32 (-1 + z) z^2 (-1 + y + z)^2 (-11 y^4\nnn & - 5 (-1 + z) z^2 +  y^3 (6 + 17 z) + y^2 (5 - 27 z + z^2)
+ y z (12 - 6 z + 5 z^2)) H(0; z)  H(2; y)\nnn & - 32 z^2 (-1 + y + z)^2 (11 y^4 (-1 + z) + 5 (-1 + z)^2 z^2 +  y^3 (8 + 7 z - 17 z^2) + y z (12
- 18 z + 11 z^2 - 5 z^3) -  y^2 (-5 + 30 z \nnn & - 28 z^2 + z^3)) H(1; z) H(3; y) +  32 y^2 (5 y^6 (-1 + z) + y^5 (20 - 31 z + 11 z^2) +  y^4 (-30
+ 75 z - 95 z^2 + 52 z^3) +  y^3 (20 - 89 z \nnn & + 209 z^2 - 188 z^3  + 48 z^4) +  y z (-12 + 62 z - 65 z^2 + 33 z^3 - 33 z^4 + 15 z^5) +  y^2 (-5
+ 52 z - 182 z^2 + 201 z^3 - 85 z^4 + 17 z^5) \nnn & -  5 (z^2 - 4 z^5 + 3 z^6)) H(0, 1; z) - 32 y^2 (-1 + y + z)^2  (5 y^4 (-1 + z) + y^3 (10
- 11 z + z^2) + z^2 (-5 - 6 z + 11 z^2)\nnn & -  y z (12 - 32 z + 11 z^2 + 11 z^3) + y^2 (-5 + 18 z - 28 z^2 + 17 z^3))  H(0, 2; y)
+ 32 y^2 (5 y^6 (-1 + z) + y^5 (20 - 31 z + 11 z^2) \nnn &+  (-1 + z)^2 z^2 (-5 - 4 z + 39 z^2) - y^4 (30 - 75 z + 41 z^2 + 2 z^3) +  y^3 (20
- 89 z + 107 z^2 - 44 z^3 + 2 z^4) -  y^2 (5 - 52 z + 140 z^2 \nnn & - 133 z^3 + 9 z^4 + 29 z^5) -  y z (12 - 68 z + 93 z^2 + 35 z^3 - 111 z^4 
+ 39 z^5)) H(1, 0; y) \nnn & -  32 y^2 (-1 + y + z)^2 (5 y^4 (-1 + z) + y^3 (10 - 11 z + z^2) +  z^2 (-5 - 6 z + 11 z^2)  - y z (12 - 32 z + 11 z^2
+ 11 z^3) \nnn &+  y^2 (-5 + 18 z - 28 z^2 + 17 z^3)) (H(0; y) H(1; z) - H(1; z) H(3; y) +  H(2, 0; y)) + 32 (-1 + y + z)^2 (5 y^6 (-1 + z) \nnn &- 5 (-1
+ z)^2 z^4 +  y^5 (10 - 11 z + z^2) + y^2 z^2 (-10 + 24 z - 17 z^2 + z^3) +  6 y^3 z (-2 + 4 z - 3 z^2 + z^3) + y z^3 (-12 + 18 z  \nnn &- 11 z^2
+ 5 z^3) +  y^4 (-5 + 18 z - 17 z^2 + 6 z^3)) H(3, 2; y)
\Big) \nnn
& -16\,C_{7, q} \Big( 2 y (-1 + z) (5 y^4 + y (10 - 19 z) z - 3 (-1 + z) z^2 +  y^3 (-10 + 21 z) + y^2 (5 - 31 z + 16 z^2)) H(0; y)\nnn & +  (-1
+ y) (2 z (-3 y^3 + 5 (-1 + z)^2 z^2 + y^2 (3 - 19 z + 16 z^2) +  y z (10 - 31 z + 21 z^2)) H(0; z) +  (-1 + z) (10 y^4 \nnn &+ 10 (-1 + z)^2 z^2
+ y^3 (-20 + 39 z) +  2 y^2 (5 - 37 z + 32 z^2) \nnn & + y z (32 - 74 z + 39 z^2))  (H(1; z) + H(2; y)))\Big)
\nnn
& -4 \,C_{6, q}\Big(8 + 391 y^2 + 112 z + 391 z^2 - 16 y (-7 + 15 z)\Big)\Bigg\}.
\end{align}
}
{\tiny
\begin{align}
S_q^ {(2),\,n_f/N}&=
 \bold{\frac{n_f}{N}} \Bigg\{  \frac{64 \,C_{1, q}}{9}  \Big(119 \zeta_{3} - 12 H(0; y)^2 H(0; z) + 12 \zeta_{2} H(1; y) +  33 \zeta_{2} H(1; z) + 21 \zeta_{2} H(2; y)
- 12 H(2; y) H(0, 1; z) \nnn & +  12 H(1; z) H(1, 0; y) + 12 H(2; y) H(1, 0; y) - 18 H(0, 1, 0; y) -  18 H(0, 1, 0; z) - 12 H(0, 1, 1; z)
- 24 H(1, 0, 0; y)\nnn & -  24 H(1, 0, 0; z) + 12 H(1, 1, 0; y)  + 24 H(1, 1, 0; z) -  6 H(0; y) (\zeta_{2} + 2 H(0; z)^2 - H(0, 1; z)\nnn &
-  H(0; z) H(1; z) ) -  6 H(0; z) (\zeta_{2} - H(1; z)^2 - 2 H(0, 2; y) + H(1, 0; y) -  2 H(2, 0; y)
- 2 H(1; z) H(2; y)) 
\Big) \nnn
 & +  \frac{64 \, C_{1, q} }{3} \Big( 33 \zeta_{2} + 26 H(0; y) H(0; z) + 26 H(0; z) H(1; z) -  26 H(0, 1; z) + 26 H(1, 0; y)\Big) \nnn
& -{64\over27}\,C_{9, q}\Big( 9 (-1 + z) (5 y^4 - y^3 (2 + z) - z^2 (1 + 7 z) -  y^2 (3 - 5 z + z^2) + y z (-4 + 5 z^2)) H(0; y) \nnn & +  (-1 + y) (9 (y^3 (-7
+ 5 z) - y^2 (1 + z^2) - y z (4 - 5 z + z^2) +  z^2 (-3 - 2 z + 5 z^2)) H(0; z) - (-1 + z)  (173 y^3 \nnn &+ y z (36 + 137 z) + y^2 (180 + 137 z)
+ z^2 (180 + 173 z))  (H(1; z) + H(2; y)))\Big)
\nnn
& -{16\over81} \,C_{6, q} \Big(2521 y^2 + 1728 y (-1 + 2 z) + z (-1728 + 2521 z)\Big)\Bigg\}.
\end{align}
}
 {\tiny
\begin{align}
S_q^ {(2),\,n_f N}&=
 \bold{N\,n_f} \Bigg\{ \frac{-64 \,C_{1, q}}{9} \Big( 12 H(0; y)^2 H(1; z) + 12 H(0; z)^2 H(2; y) +  3 H(0; y) (3 \zeta_{2} + 2 H(0; z) H(1; z) - 10 H(1; z)^2
- 10 H(2; y)^2 \nnn &-  10 H(0, 1; z)  + 8 H(0, 2; y) - 4 H(3, 2; y) ) + 3 H(0; z) (3 \zeta_{2} + 8 H(1; z)^2
-  10 H(2; y)^2 - 4 H(1; z) (2 H(2; y) + H(3; y)) \nnn &  - 4 H(0, 2; y) -  2 H(1, 0; y) + 4 H(2, 0; y) - 4 H(3, 2; y) )
+ 2 (-43 \zeta_{3} + 6 \zeta_{2} H(1; y) +  30 H(1; z)^2 H(3; y) + 30 H(2; y) H(0, 1; z)\nnn & +  12 H(3; y) H(0, 1; z) + 6 H(2; y) H(1, 0; y)
+ 12 H(0, 0, 1; z) -  12 H(0, 0, 2; y) - 9 H(0, 1, 0; y) - 3 H(0, 1, 0; z) +  6 H(0, 1, 1; z)\nnn & - 12 H(1, 0, 0; y) + 6 H(1, 1, 0; y)
-  24 H(1, 1, 0; z) + 12 H(2, 0, 0; y) + 12 H(2, 1, 0; y)  +  48 H(2, 3, 2; y) + 60 H(3, 2, 2; y)\nnn & + 12 H(3, 3, 2; y)
   +  6 H(1; z) (\zeta_{2} + 8 H(2; y) H(3; y) + H(3; y)^2\nnn & - 5 H(0, 2; y) -  H(0, 3; y) + H(1, 0; y) - 5 H(2, 0; y) - H(3, 0; y) +  2 H(3, 2; y)))
\Big) \nnn
& + C_{5, q} \Big((-64 z^2 (-25 y^4 + 36 y (-1 + z)^2 z + 18 (-1 + z)^2 z^2 +  6 y^3 (1 + 3 z) + y^2 (12 - 84 z + 47 z^2)) H(0; z) H(2; y))/9\nnn &
+  (160 y^2 z^2 (y^2 + z^2) (\zeta_{2} + (51 H(0; y)^2)/5 + 8 H(0; y) H(0; z) +  (51 H(0; z)^2)/5 - (106 H(0; z) H(1; z))/5
+ (174 H(1; z)^2)/5 \nnn &+  (348 H(1; z) H(2; y))/5 + (174 H(2; y)^2)/5))/9 +  (128 y^2 (3 y^4 + 6 y^3 (-1 + z) + z^2 (2 + z - 13 z^2) -  y^2 (-3
+ 12 z + z^2)\nnn & + y z (6 - 14 z + 3 z^2))  (H(0, 1; z) + H(1, 0; y)))/3 -  (64 y^2 (18 y^4 + 36 y^3 (-1 + z) + z^2 (12 + 6 z - 25 z^2)
+  6 y z (6 - 14 z + 3 z^2)\nnn & + y^2 (18 - 72 z + 47 z^2))  (H(0; y) H(1; z) + H(0, 2; y) + H(2, 0; y)))/9 +  (128 (3 y^6 + 6 y^5 (-1 + z)
+ 6 y (-1 + z)^2 z^3 \nnn &+ 3 (-1 + z)^2 z^4 +  y^2 z^2 (4 - 13 z - 14 z^2) + y^4 (3 - 12 z - 14 z^2) +  y^3 z (6 - 13 z
+ 6 z^2)) (H(1; z) H(3; y) + H(3, 2; y)))/3
\Big) \nnn
& -{64\over27} C_{8, q}\Big( y (-1 + z) (54 y^5 + (27 - 490 z) z^3 + 4 y^4 (-27 + 163 z) +  y z^2 (108 - 643 z + 508 z^2)\nnn & + y^2 z (135 - 832 z + 634 z^2)
+  y^3 (54 - 787 z + 724 z^2)) H(0; y) +  (-1 + y) (z (54 (-1 + z)^2 z^3 + y^4 (-490 + 508 z)\nnn & +  4 y^2 z (27 - 208 z  + 181 z^2) + y^3 (27 
- 643 z + 634 z^2) +  y z^2 (135 - 787 z + 652 z^2)) H(0; z) +  3 (-1 + z) (18 y^5 \nnn & + 18 (-1 + z)^2 z^3 - y^4 (36 + 497 z) +  y z^2 (78 - 66 z
- 497 z^2) + y^2 z (78 - 288 z - 383 z^2)\nnn & +  y^3 (18 - 66 z - 383 z^2)) (H(1; z) + H(2; y)))\Big)
\nnn
& + {16\over81} C_{6, q}  \Big(36557 y^2 - 3024 y (-1 + 2 z) + z (3024 + 36557 z)\Big) \Bigg\}.
\end{align}
}
 {\tiny
\begin{align} 
S_q^ {(2),\, n_f^2}&=
 \bold{n_f^2} \Bigg\{ \frac{-32 \, C_{1, q} }{9}  \Big( 3 H(0; y)^2 + 3 H(0; z)^2 +  2 H(0; y) (H(0; z) - 4 H(1; z)) - 8 H(0; z) (H(1; z) + H(2; y)) +  2 (\zeta_{2}\nnn &
+ 12 H(1; z)^2 + 24 H(1; z) H(2; y) + 12 H(2; y)^2 -  4 H(0, 2; y) - 4 H(2, 0; y))
\Big)\nnn
& + \frac{ 640 \, C_{1, q}}{9}  \Big(H(0; y) + H(0; z) - 4 (H(1; z) + H(2; y))  
\Big)\nnn
 & -\frac{6400\, C_{1, q}} {27} \Bigg\}.
\end{align}
}
where the constants $C_{i,q}$ are
{\tiny
\begin{eqnarray}
C_{1, q} &=& \frac{Q^2 (y^2 + z^2)}{(-1 + y+ z)}\,,C_{2, q} = \frac{Q^2}  {(-1+y)^2 (-1+z)^2(-1+y+z)}\, ,
C_{3, q} = \frac{Q^2} {y^2 z^2   (-1+y+z) ^2 (-1+y) (-1+z) },\nnn
C_{4, q} &=&  \frac{Q^2}{ y z (-1+y+z)} \,,C_{5, q} = \frac{Q^2} {y^2 z^2 (-1+y+z)}\,, C_{6, q} = \frac{Q^2}{(-1 + y+ z)}\,,C_{7, q} = \frac{Q^2}{yz(-1+y)(-1+z)(-1+y+z)},\nnn
C_{8, q} &=& \frac{Q^2}{yz(-1+y)(-1+z)(-1+y+z)(y+z)}\,,C_{9, q} = \frac{Q^2}{(-1+y)(-1+z)(-1+y+z)(y+z)}\,.
\end{eqnarray}
}
\bibliographystyle{JHEP}
\bibliography{main}
\end{document}